\documentclass[twocolumn,fleqn]{cas-dc}

\usepackage{amsmath}
\usepackage{graphicx,epstopdf}
\usepackage{array}
\usepackage{dcolumn}
\usepackage{hepparticles}
\usepackage{heppennames}
\usepackage{color}
\usepackage{blindtext}
\usepackage{booktabs}

\newcounter{gg}

\newcommand{\tRC}{\tau_{\scalebox{0.5}{RC}}}
\newcommand{\tLR}{\tau_{\scalebox{0.5}{LR}}}
\newcommand{\tRCs}{\tau_{\scalebox{0.4}{RC}}}

\DeclareRobustCommand{\Hbar}{\HepAntiParticle{H}{}{}\xspace}

\DeclareRobustCommand{\pbar}{\HepAntiParticle{p}{}{}\xspace}

\DeclareRobustCommand{\pos}{\HepParticle{e}{}{+}\xspace}

\usepackage[numbers,square]{natbib}

\begin{document}

\title[mode=title]{Two-Symmetry Penning-Ioffe Trap for Antihydrogen Cooling and Spectroscopy}

\shorttitle{Penning-Ioffe Trap for Antihydrogen Studies}
\shortauthors{E. Tardiff et~al.}


\author[1]{E. Tardiff}[type=editor,
                       orcid=0000-0002-5245-1490]

\address[1]{Center for Fundamental Physics, Northwestern University, Evanston, IL 60208, USA}


\author[1,2]{X. Fan}

\address[2]{Department of Physics, Harvard University, Cambridge, MA 02138, USA}


\author[1]{G. Gabrielse}
\ead{gerald.gabrielse@northwestern.edu}


\author[3]{D. Grzonka}

\address[3]{IKP, Forschungszentrum J{\"u}lich, 52425 J\"ulich, Germany}


\author[2]{C. Hamley}


\author[4]{E.A. Hessels}

\address[4]{York University, Department of Physics and Astronomy, Toronto, ON M3J 1P3, Canada}


\author[2]{N. Jones}


\author[5]{G. Khatri}

\address[5]{CERN, 1211 Geneva 23, Switzerland}


\author[6]{S. Kolthammer}

\address[6]{Department of Physics, Imperial College, London SW7 2AZ, UK}


\author[1]{D. {Martinez Zambrano}}


\author[2]{C. Meisenhelder}


\author[2]{T. Morrison}


\author[2]{E. Nottet}


\author[7]{E. Novitski}

\address[7]{Department of Physics, University of Washington, Seattle, WA 98195, USA}



\author[4]{C.H. Storry}

\date{\today}

\begin{abstract}     
High-accuracy spectroscopic comparisons of trapped antihydrogen atoms (\Hbar) and hydrogen atoms (H) promise to stringently test the fundamental CPT symmetry invariance of the standard model of particle physics.  ATRAP's nested Penning-Ioffe trap was developed for such studies.  The first of its unique features is that its magnetic Ioffe trap for \Hbar atoms can be switched between quadrupole and octupole symmetries.  The second is that it allows laser and microwave access perpendicular to the central axis of the traps.  
\end{abstract}


\maketitle

\tableofcontents



\section{Introduction}

Low energy antiproton (\pbar) and antihydrogen (\Hbar) physics began with the 1986 demonstration that \pbar beams could be slowed to and trapped at \cite{PbarCapture}  energies ten orders of magnitude lower than previously realized, using CERN's Low Energy Antiproton Ring (LEAR). At the time, the proposed use of such cold \pbar for low energy \Hbar production \cite{NestedPenningTrap} and the magnetic trapping of \Hbar atoms \cite{Erice} were radical departures from the beam-based approaches previously envisioned \cite{HerrAntihydrogen}.   

The nested Penning-Ioffe traps proposed for synthesizing \Hbar atoms \cite{Erice} have since demonstrated their worth by producing and capturing all the trapped \Hbar atoms realized so far. A nested Penning-Ioffe trap \cite{NestedPenningTrap} is the superposition of a Penning trap  (used to trap \pos and \pbar) with a magnetic-field-minimum trap \cite{Ioffe} (to trap the neutral \Hbar produced from \pos and \pbar). The basic ideas of a nested Penning trap and of a Ioffe trap are briefly summarized in Sections~\ref{sec:NestedTrap} and \ref{sec:IoffeTrap}, respectively.  A nested Penning-Ioffe trap is briefly summarized in Sec.~\ref{sec:NestedPenningIoffeTrap}.   

The first Penning-Ioffe trap was built by ATRAP.  It used a quadrupole Ioffe field \cite{JuelichIoffeTrap} and was used to demonstrate that \Hbar atoms could be produced within these fields \cite{AntihydrogenProducedInPenningIoffeTrap} and to confine ground-state \Hbar atoms  \cite{AtrapTrappedAntihydrogen}. This Penning-Ioffe trap is discussed in Sec.~\ref{sec:FirstPenningIoffeTrap} for the first time partly because of its historical importance, but primarily to contrast design and construction methods that are very different from those used for the trap we focus on in this work.     
 
Meanwhile, others used an octupole Penning-Ioffe trap \cite{AlphaMagneticTrap} to confine \Hbar atoms \cite{AlphaTrappedAntihydrogen2010}. A third trap, ATRAP's second-generation trap, differs in that it can produce either a quadrupole field, an octupole field, or a combination.  Recent simulations carried out for laser cooling of trapped \Hbar atoms and for spectroscopy of \Hbar atoms, soon to be reported, will establish the relative advantages and disadvantages of the two symmetries. 

The mentioned \Hbar traps have similar depths (within a factor of 2), limited by the critical fields and currents of their superconducting windings, their size, and by how close the trapped \Hbar can get to these windings. The first Penning-Ioffe trap required much less current than its successors because a large number of windings were used. Not much liquid helium was required to keep this trap cold because the smaller current could be sent into its dewar through smaller current leads that brought less heat into the dewar. However, the high inductance required fifteen minutes to energize the trap, and, as designed, ten minutes to remove its current to turn the trap off.  

The second and third Penning-Ioffe traps were low inductance traps that used many fewer Ioffe trap windings. Their coils' lower inductances allowed them to be turned on in seconds and off in milliseconds, greatly reducing the difficulty of distinguishing \Hbar annihilation signals from cosmic ray signals during the time that \Hbar exit the trap as it is turned off. These Ioffe traps require a much higher current, and a great deal more liquid helium. 

The focus of this report is ATRAP's design and construction of its second-generation trap (Sec.~\ref{sec:CTRAP}), and a demonstrated performance comparable to design expectations (Sec.~\ref{sec:Operation}).  The biggest challenge came from the unique choice to include radial sideports to allow laser beams and microwaves to enter the Penning-Ioffe traps perpendicular to the central symmetry axis of the traps. The choice to be able to produce a Ioffe trap of either a quadrupole or octupole symmetry also added to the challenge.

We illustrate the usefulness of the low-inductance, two-symmetry, Penning-Ioffe trap with two examples (Sec.~\ref{sec:Usefulness}).  The first example is images of ejected plasmas that were captured, cooled and rotated to achieve a small diameter within the Penning trap (Sec.~\ref{sec:PositronPlasmas}).  The second example consists of signals from \Hbar atoms that were confined in the Ioffe trap until it was rapidly turned off (Sec.~\ref{sec:TrappedAntihydrogen}).



\section{Superimposed Traps}

\subsection{Nested Penning Traps for \Hbar Production}
\label{sec:NestedTrap}

\begin{figure}
	\centering
	\includegraphics*[width=0.8\columnwidth]{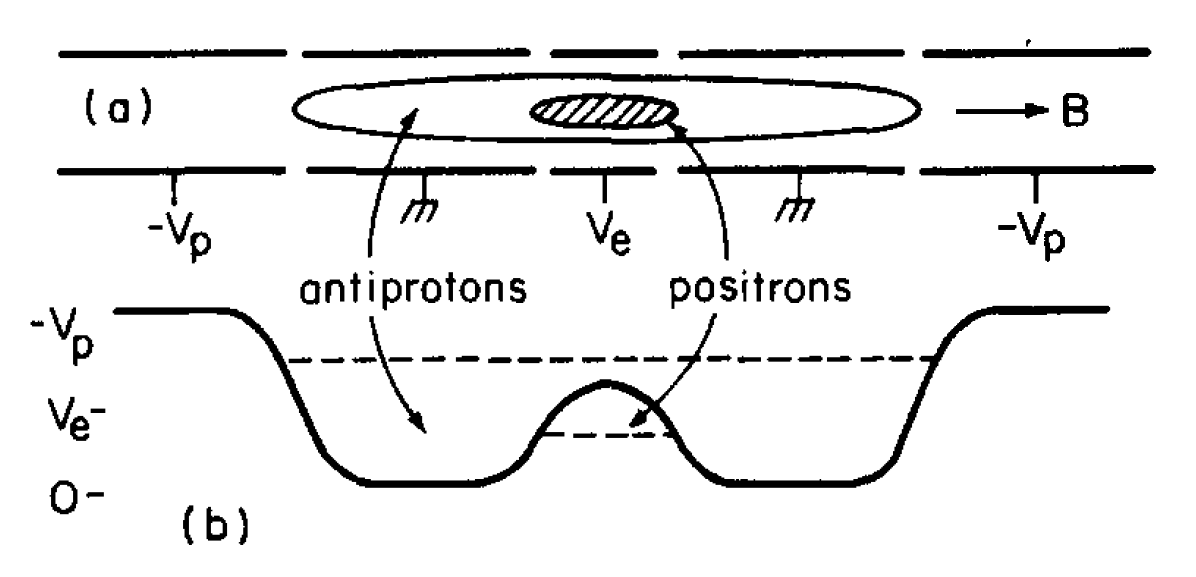}
	\caption{A cross section of a cylindrically symmetric nested Penning trap (a) and the potentials   that produce trap wells for \pbar and \pos (b).  Taken from \cite{NestedPenningTrap}.}  
	\label{fig:NestedPenningTrap}
\end{figure}

The nested Penning trap was invented \cite{NestedPenningTrap} to bring cold, oppositely-charged \pbar and \pos into contact long enough to produce cold \Hbar. All of the trapped \Hbar so far has been produced in such a device. A nested Penning trap starts with a uniform magnetic field along the symmetry axis of the trap electrodes,  
\begin{equation}
\vec{\bf B} = B_0\hat{\bf z}.
\label{eq:BiasField}
\end{equation} 
Opposite sign charges are suspended in the outer and inner wells of a cylindrically symmetric, nested Penning trap (Fig.~\ref{fig:NestedPenningTrap}a) that produces potential wells (Fig.~\ref{fig:NestedPenningTrap}b). Negative \pbar in the outer well, when given enough energy to go over the central potential barrier, interact with positive \pos to form \Hbar atoms mainly through three-body collisions. 

The nested Penning trap was initially demonstrated with protons and electrons \cite{ElectronCoolingOfProtons}. 
During the last week of LEAR's operation in 1999, \pbar and \pos were confined in a nested Penning trap for the first time \cite{Ingredients}. The \pos observably cooled the \pbar, demonstrating there was an interaction as needed to form \Hbar atoms.  Cold \pbar and \Hbar physics continued at CERN's Antiproton Decelerator (AD), with evidence of \Hbar formation in a nested Penning trap reported in 2002 \cite{2002AthenaNatureLetter,2002AtrapBackgroundFreeAntihydrogen,2002AtrapDrivenAntihydrogenProduction}.

The strength of the magnetic field determines how rapidly the charged particles cool via synchrotron radiation.  Due to the \pbar mass, the radiation rate is too slow to be useful.  In the 6 T field often used for precision measurements and for initial \Hbar experiments, however, a \pos radiates into free space with a damping time of 0.1 s.  As discussed in more detail below, the constraints introduced by superimposing a magnetic-minimum Ioffe trap on the Penning trap mean that the uniform field should not exceed about 1 T.  In this field, the \pos radiation time (going as $|B|^{-2}$) of 4 s increases the time it takes to manipulate \pbar and \pos to form cold \Hbar atoms.

\subsection{Ioffe Traps for \Hbar Confinement}
\label{sec:IoffeTrap}

Atoms in low-magnetic-field seeking states can be confined near a magnetic field minimum.  A Ioffe trap \cite{Ioffe} (sometimes called a Pritchard-Ioffe trap \cite{PritchardProposesIoffeTrap}) is a practical way to generate a confining field while maintaining a magnetic field at the minimum suitable for a Penning trap.  
  
\begin{figure}
	\centering
	\includegraphics*[width=\columnwidth]{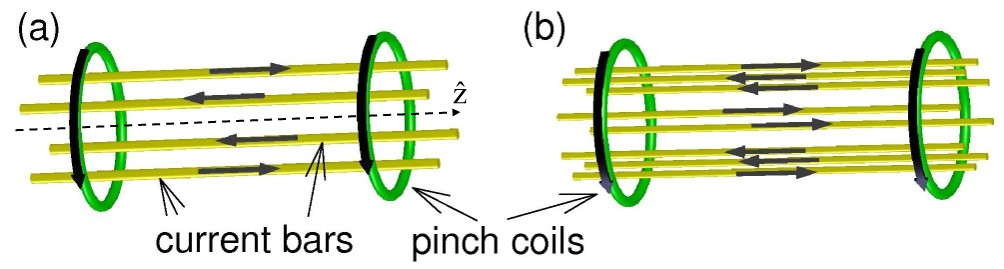}
	\caption{The magnetic minimum at the center of quadrupole (a) and octupole (b) Ioffe traps are produced by $2n=4$ and $2n=8$ vertical current bars, respectively, plus orthogonal pinch coils.}  
	\label{fig:QuadrupoleAndOctupole}
\end{figure}

 An idealized Ioffe trap (Fig.~\ref{fig:QuadrupoleAndOctupole}) is composed of two elements: 2$n$ current wires running parallel to, and spaced circularly around, the z-axis, and two current loops centered on the z axis and separated by a distance large compared to their diameters.  Current runs in opposite directions in neighboring wires and in the same direction in the two loops, creating a magnetic minimum at the center.  The total field can be written as $\vec{\bf B} = B_z \hat{\bf z} + \vec{\bf B}_\perp$. The transverse magnetic field produced near the center of the trap in the $z=0$ midplane (in terms of cylindrical coordinates, $\rho$ and $\phi$) is 
 \begin{equation}
 \vec{\bf B}_\perp(\rho,\phi,z\approx 0) \approx B_R \left(  \frac{\rho}{R} \right)^{n-1} \left(cos(n\phi) \hat{\mathbf{\rho}} - sin(n\phi) \hat{\bf \phi}\right).
 \label{eq:IoffeB}
 \end{equation}
 A magnetic minimum is produced when $2n= 4, 6, 8, ...$.  The constant $B_R$ is the field magnitude that corresponds to a radius $R$ within the trap windings.   Fig.~\ref{fig:QuadrupoleAndOctupole} illustrates for a quadrupole with $2n=4$ in (a), and for an octupole with $2n=8$ in (b).

 The magnetic moment of a ground state \Hbar comes primarily from the positron (with charge $e$ and mass $m$).  This is essentially one Bohr magneton, $\mu_B=e\hbar/(2m)$.  Near the center of the Ioffe trap, if the field from the distant pinch coils is neglected, the potential energy for a low-field-seeking, ground state \Hbar would be 
   \begin{eqnarray}
  U(\rho,\phi,z\approx 0)&=&-\mu_B |\vec{\bf B}_\perp (\rho,\phi,z\approx0)|\\
   &=& -\mu_B B_R \left( \frac{\rho}{R}    \right)^{n-1}.
  \label{eq:IoffeField} 
  \end{eqnarray} 
 For a quadrupole ($2n=4$ current bars), $U \propto \rho$.  For an octupole ($2n=8$ current bars), $U \propto \rho^3$.  A later figure (Fig.~\ref{fig:RadialWells}) contrasts more realistic radial Ioffe fields with these idealized fields. 
 
Interestingly, the motion of a trapped \Hbar is integrable within a quadrupole Ioffe trap, but not within an octupole trap.  The result is that the energy in the radial and axial motions within these traps is decoupled for a quadrupole trap, but coupled for an octupole, as we will discuss soon in a report on simulations of \Hbar motions in our traps.  This difference persists in a realistic Penning-Ioffe trap, with consequences for how many lasers are required to cool all the motions of the trapped \Hbar. Because the new trap featured in this work can produce fields of both symmetries, it allows other possibilities.  A quadrupole trap with controllable addition of an octupole component, for example, could be used to control and manipulate charged particle loss and \Hbar laser-cooling rates.

\subsection{Nested Penning-Ioffe Traps for \Hbar Production and Confinement} 
\label{sec:NestedPenningIoffeTrap}
  
Obtaining trapped \Hbar atoms requires the simultaneous application of charged and neutral particle traps \cite{Erice} for producing and then confining \Hbar. The properties of each type of trap are significantly modified by the presence of the other.  

For charged particles confined in a Penning trap field $B_0\hat{\bf z}$ and an axially symmetric trapping potential, conservation of energy and angular momentum together ensure that they remain confined indefinitely \cite{ONeilConfinementTheorem}. This changes when the Ioffe field gradient is superimposed to confine \Hbar atoms.  The Ioffe field ${\bf B}(\rho,\phi,z)$ breaks the axial symmetry of the Penning trap, so charged particle confinement is no longer guaranteed \cite{PenningIoffeTheory}.  Significant loss of \pbar and \pos can happen away from the central axis, making it desirable to apply the trapping field quickly and for only a short time to minimize these losses.  There is typically a radius outside of which the radial field of the Ioffe trap directs all charged particles into the trap walls \cite{2001AtrapFirstPositronCooling}.  An octupole field, with less field near the trap axis than a quadrupole, reduces charged particle losses \cite{FajansHigherOrderTraps} during the time required for \Hbar formation.  

For \Hbar trapping, the Ioffe gradient field is modified by the vector addition of the spatially uniform Penning trap field, so the effective trapping potential energy is 
\begin{equation}
U(\rho,\phi,z) = -\mu_B|B_0\hat{\bf z}+{\bf B}(\rho,\phi,z)|. 
\label{eq:PenningIoffeField} 
\end{equation}
The vector sum with the uniform field increases $U$ more at the trap minimum than at the perimeter, thus decreasing the overall trap depth.  Our Ioffe trap coils are designed to make the strongest possible Ioffe field gradient that does not make the superconducting coils quench. With the strongest gradient fields that can be produced, the uniform field that can be added without significantly reducing the depth of the neutral particle trap is the $B_0 = 1$ T that we use. 

A realistic Penning-Ioffe trap introduces other variations that are difficult to calculate analytically. Infinitely extended current bars (e.g. Fig.~\ref{fig:QuadrupoleAndOctupole}) are not possible, of course. The same current is instead sent through all the current bars by connecting them to make ``racetrack'' coils (illustrated in Fig.~\ref{fig:IoffeTrapWindings}.) These connections add to or subtract from the pinch coil field depending on the relative direction of the currents.   Realistic pinch coils also cannot be far from the trap center.  Finally, providing sideports for radial access makes it necessary to spread the octupole current bars in 4 locations near the trap center.  

When \Hbar trapping was reported in a quadrupole Penning Ioffe trap \cite{AtrapTrappedAntihydrogen} it was assumed that the realistic geometry for a quadrupole trap described above would make the particle trajectories within this trap non-integrable.  This assumption was questioned based upon model simulations \cite{Zhmoginov2013}.  To check, we calculated and will soon publish \Hbar trajectories for the strongest quadrupole Penning-Ioffe trap we have realized for our second generation trap.  The trajectories remain remarkably integrable, despite the field modification due to the current return paths of the racetracks.  It remains to be studied how adding small octupole fields changes this axial-radial coupling time.



\section{First Penning-Ioffe Trap}
\label{sec:FirstPenningIoffeTrap}

The ATRAP first generation Penning-Ioffe trap was used for a number of experiments ranging from producing \Hbar atoms within such fields \cite{AntihydrogenProducedInPenningIoffeTrap}, to confining 5 trapped \Hbar atoms per trial on average \cite{AtrapTrappedAntihydrogen}.  The design and construction of this trap, detailed in Ref.~\cite{JuelichIoffeTrap}, are summarized here because of its initial importance, and to contrast it to the second generation trap which is the focus of the rest of this report.

The biggest challenge to the ATRAP Penning-Ioffe traps comes from the feature that distinguishes our traps from others.  This is the choice to have sideports that allow laser, microwave and atom beams to travel between the superconducting Ioffe windings and then through the electrodes of the Penning-Ioffe trap, traveling perpendicular to its central axis. The intent was to facilitate laser cooling along multiple axes (essential in a quadrupole Ioffe trap) along with providing more laser and microwave spectroscopy options. 

\begin{figure}
	\centering
	\includegraphics*[width=\columnwidth]{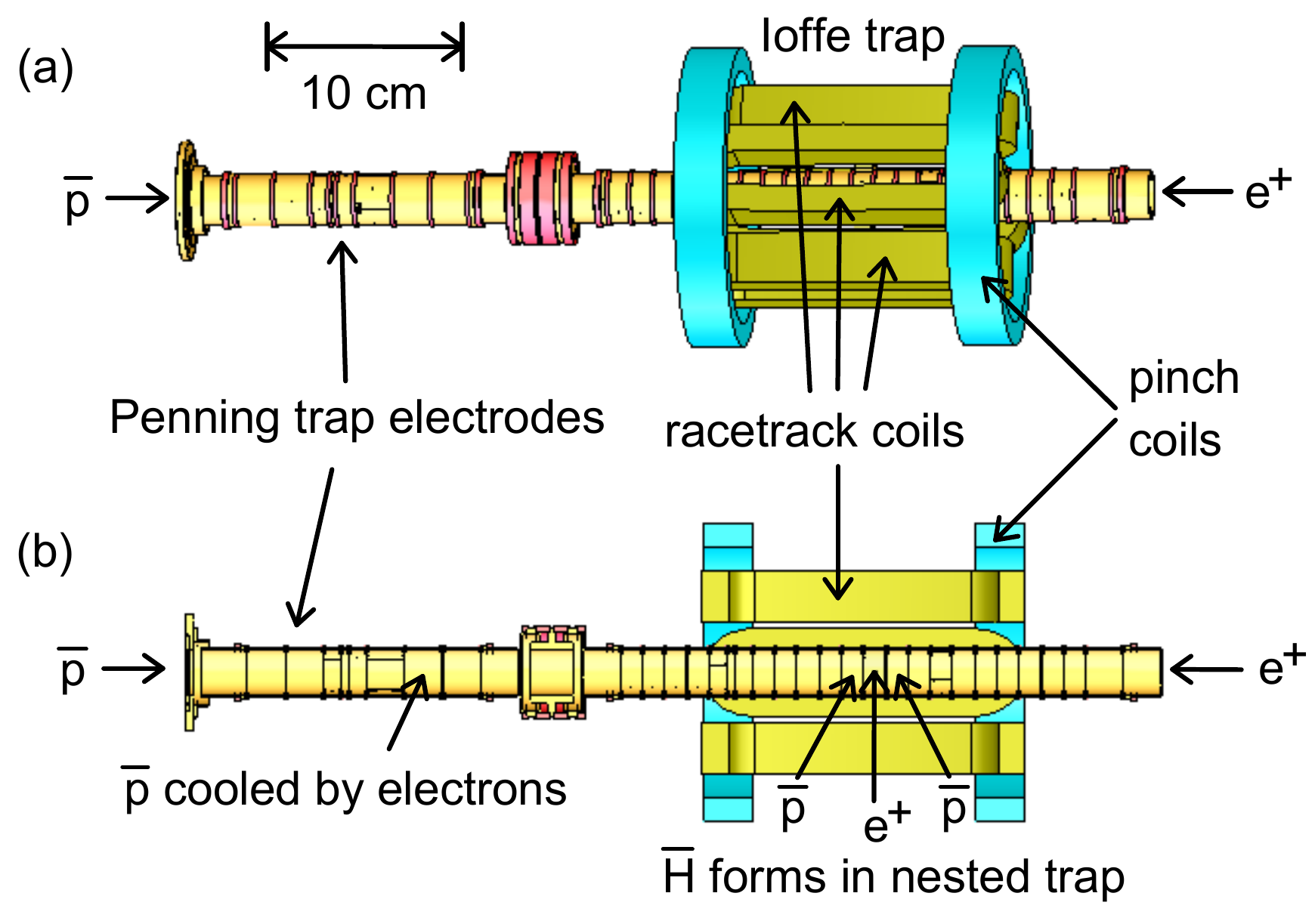}
	\caption{A scale representation of the ATRAP first generation Penning-Ioffe trap.  There is a spatially uniform magnetic field directed parallel to the trap axis that is produced by an external solenoid (not shown). }
	\label{fig:PenningIoffeTrapGenerationOne}
\end{figure}

\begin{figure}
	\centering
	 \includegraphics[clip=true,width=\columnwidth]{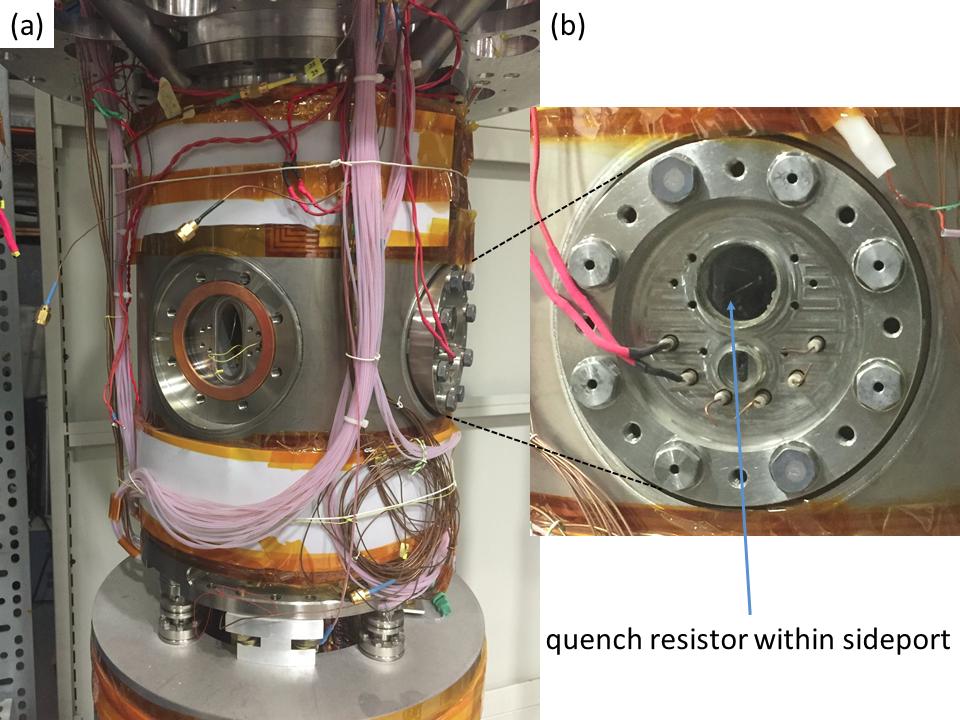}
		\caption{Generation 1 Ioffe trap enclosure (a), with 2 of 4 side ports showing, and a \pbar loading solenoid below.  A quench resistor attached to one side port is visible through the window in (b).  Typically 10 W would cause a quench within 1 sec.}
	\label{fig:Btrap}
\end{figure}

\begin{figure}
	\centering
		 \includegraphics[clip=true,width=\columnwidth]{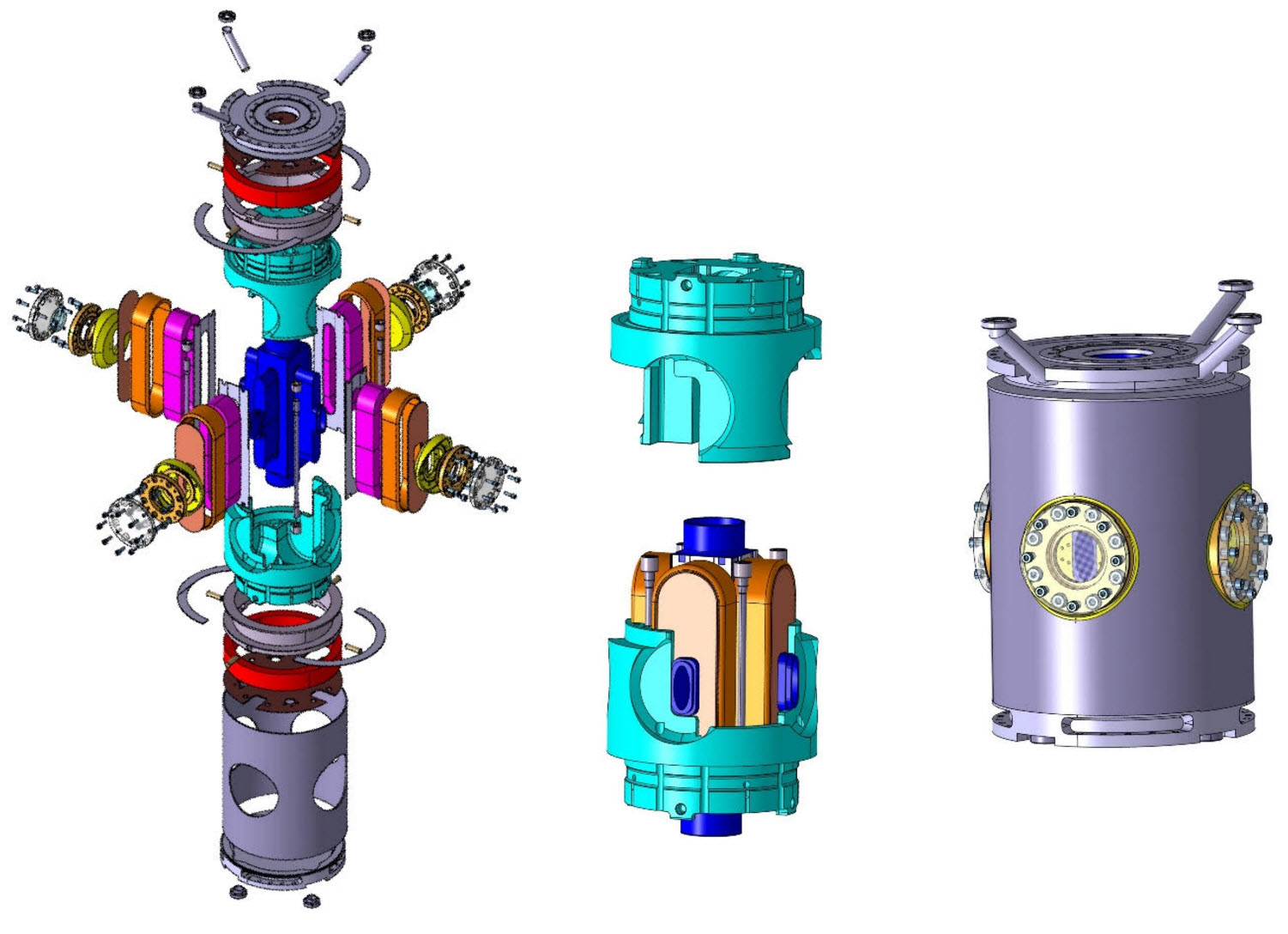}
	\caption{Exploded view of the first generation Ioffe trap.}
	\label{fig:IoffeTrapOneExplodedView}
\end{figure}

This first Penning-Ioffe trap (Fig.~\ref{fig:PenningIoffeTrapGenerationOne}) utilized high-inductance Ioffe coils to minimize the current needed to produce the deepest practical \Hbar trap.  For comparison to the second generation trap to be discussed, Fig.~\ref{fig:Btrap} shows pictures of this first generation trap and Fig.~\ref{fig:IoffeTrapOneExplodedView} represents an exploded mechanical view. A 1 Tesla bias field for the Penning trap was produced by a persistent superconducting solenoid (outside the view of the figure) that stayed on continuously for many months at a time.

The Ioffe racetrack coils (2930 turns in 40 layers) and pinch coils (2558 turns in 36 layers each) were wound on titanium forms.  The superconducting wire is type F54-1.35 from EAS (European Advanced Superconductors).  These wires have a copper diameter of 0.54 mm within which 54 NbTi filaments are embedded.  

The windings were stabilized with aluminum clamps that contain the considerable stress produced within the energized coils.  Detailed stress calculations characterized the shrinking during cooling and the strong Lorentz forces.  To ensure a well defined pre-stress at the coils, final clamp dimensions were based upon coil dimensions that were measured to a precision of $\pm 50~ \mu$m.

The exploded drawing in Fig.~\ref{fig:IoffeTrapOneExplodedView} shows the separate components of the system in the titanium housing which was welded to the coil form and the end plates after assembling. The components were manufactured at the Research Center Forschungszentrum J\"ulich (FZJ).  The coil winding, assembly and some of the  welding were carried out by ACCEL.

The large number of windings made it possible to use modest currents (compared to our second generation trap) to establish a magnetic trap.  Just 80 A in each pinch coil and 68 A in the racetrack coil produced the deepest trap possible without quenching the superconducting windings.  This trap was 0.56 Tesla deep when a 1 Tesla bias field was applied along the central axis of the trap.  This corresponds to a 380 mK trap depth for an \Hbar atom in a low-field-seeking ground state.  

The high inductance due to the large number of windings caused significant experimental limitations.  It took nearly 15 minutes to ramp the current up to the full field without triggering diodes that were installed across the coils to protect them from damage if their superconducting windings quenched. During this time the magnetic gradient caused substantial losses of \pos and \pbar stored in the nested Penning trap.  
The large inductance, along with quench protection diodes permanently attached across each coil, also made it impossible to ramp down the current faster than ten minutes.  This time was much longer than desired in that cosmic rays triggered our detectors many times during the time over which we released trapped \Hbar atoms. 

To get a faster turn off time, the coils were quenched deliberately in one of two ways.  Either the windings were heated with a heat pulse applied to a resistor (see Fig.~\ref{fig:Btrap}b) or the current in a coil was increased past its critical current.  Neither procedure is recommended for longevity, neither happens at a precisely controlled time, and a quench still did not turn off the trapping field in less than 1 second \cite{AtrapTrappedAntihydrogen}.  Also, it typically took the rest of an 8 hour antiproton beam shift (and sometimes longer) before the trap had recovered enough to be re-energized for another trial.    

This Penning-Ioffe trap was the first one to provide radial access to the interior of the trap for lasers and microwaves, through four sideports spaced at 90 degrees from each other (Fig.~\ref{fig:Btrap}).  The access paths fit between the windings of the 4 quadrupole current bars which are connected to form racetracks. The photos in  Fig.~\ref{fig:Btrap} show the vacuum enclosure, a sideport and the attached quench resistor.  These ports were designed to facilitate both laser cooling and spectroscopy of trapped \Hbar atoms, and also made it possible to demonstrate both the laser-controlled production of Rydberg positronium \cite{2004AtrapLaserControlledPositronium,2016AtrapLaserControlledPositronium} and \Hbar \cite{2004AtrapLaserControlledAntihydrogen}. 

For laser-controlled \Hbar production, a beam of Cs atoms from an oven located in one sideport was sent through the trap after the atoms were  excited to Rydberg states with red and green laser light sent into the 4 K environment through an optical fiber. A charge exchange collision between the Rydberg Cs and trapped positrons produced Rydberg positronium. A subsequent collision of a Rydberg positron and a trapped antiproton produced Rydberg \Hbar. This laser-controlled, two-step charge exchange production of \Hbar atoms is the only demonstrated alternative to the three-body formation of slow \Hbar \cite{2004AtrapLaserControlledAntihydrogen} in a nested Penning trap \cite{NestedPenningTrap}.



\section{Second Generation Design and Construction}
\label{sec:CTRAP}

\begin{figure}
	\centering
	\includegraphics*[width=\columnwidth]{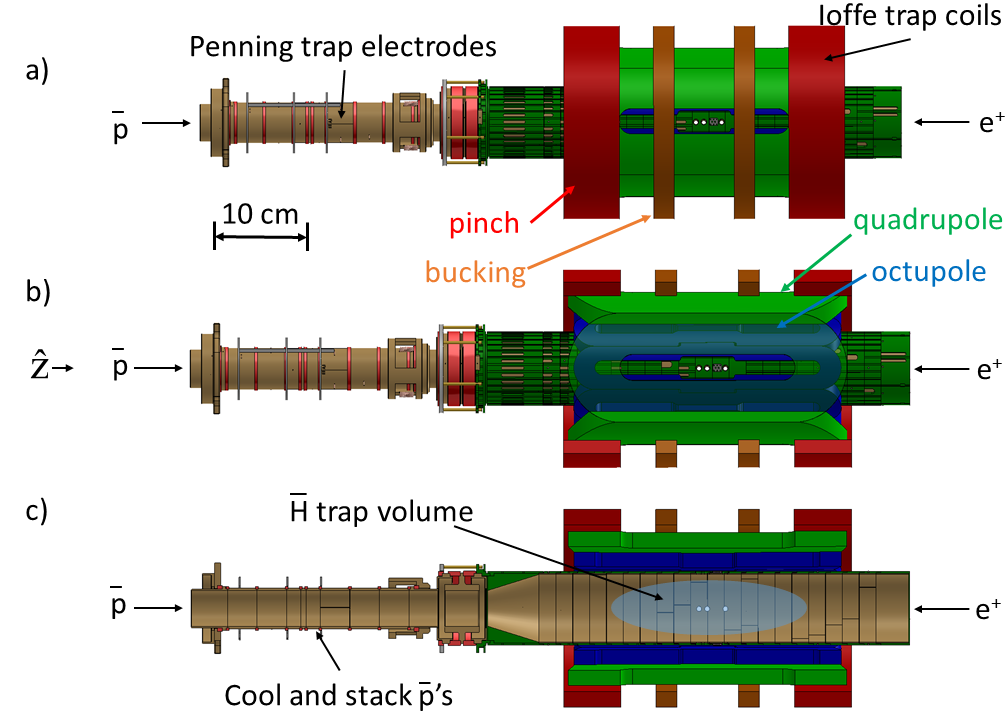}
	\caption{Three scaled representations of the electrodes and coils of ATRAP's latest nested Penning-Ioffe trap.  A spatially uniform magnetic field, directed parallel to the trap axis, is produced by an external solenoid outside the view of this figure.  The central axis of the trap is defined as the $\hat{\textrm{z}}$ axis, oriented to the right in the figure and upwards in the installed aparatus.}
	\label{fig:NestedPenningIoffeGenerationTwo}
\end{figure}

The positioning of the components of ATRAP's second-generation Penning-Ioffe trap are similar to the first, as illustrated by comparing Figs.~\ref{fig:PenningIoffeTrapGenerationOne} and \ref{fig:NestedPenningIoffeGenerationTwo}.  The section of the Penning trap electrodes devoted to cooling and stacking \pbar has identical dimensions to the first-generation, while the electrode design for the \Hbar trapping volume was altered to optimize the \Hbar trap depth given the field profile generated by the Ioffe coils. The field profiles are discussed in more detail in Sec.~\ref{sec:HbarPotentials}.

\subsection{Ioffe Coils}
\label{sec:IoffeCoilSection}

\begin{figure}
	\includegraphics*[width=3.25in]{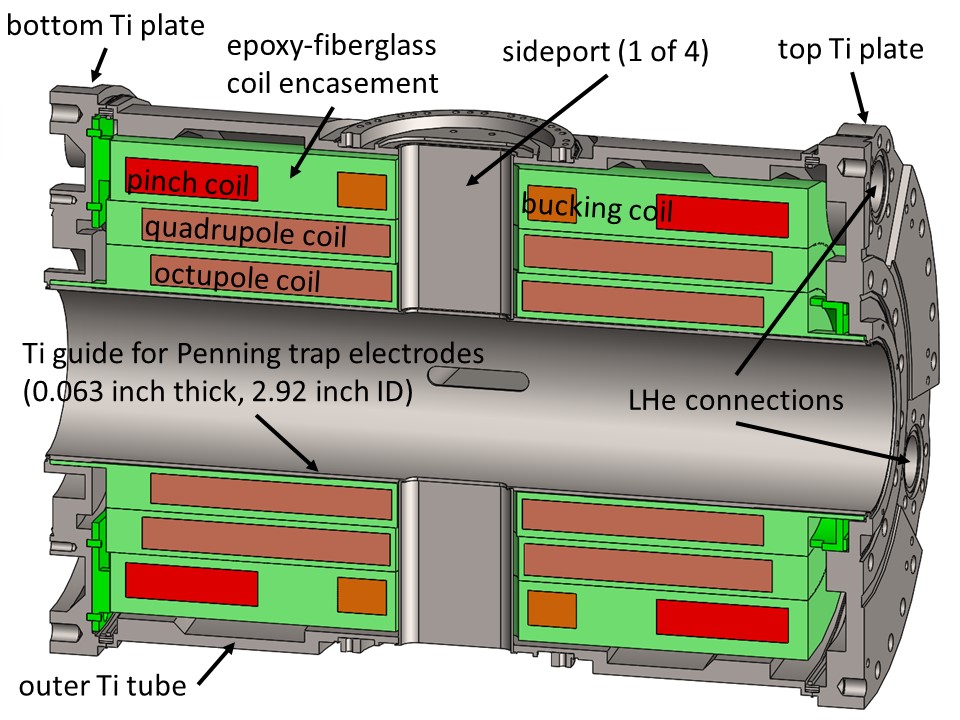}
	\caption{Scale cross section of the Ioffe trap and its enclosure.  }  
	\label{fig:IoffeTrapCrossSection}
\end{figure}

The principle design goal for the second-generation coils was to make the fastest possible charging and discharging times using very low inductance coils for the Ioffe trap, while maximizing the achievable trap depth \cite{ThesisKolthammer}.  To maintain compatibility with the existing dewar and Penning trap apparatus, many external properties of ATRAP's latest Ioffe trap are similar to those of the first. It is enclosed in a vacuum container that has the same outer dimensions as the first generation (Sec.~\ref{sec:FirstPenningIoffeTrap}), and it has similarly sized oval sideports to allow laser and potential microwave access into the center of the Penning-Ioffe trap (see Fig.~\ref{fig:IoffeTrapCrossSection}).  However, the internal Ioffe windings and the fabrication method are very different. And, this second-generation trap requires an external quench detection and current removal system to avoid the destruction of the coils. 

As in the first generation, a superconducting solenoid (not shown in the figures) surrounds the trap and applies a spatially uniform 1 Tesla magnetic field along the central axis of the trap. The persistent current in this solenoid typically stays on for many months. An additional ``loading solenoid'' surrounds the lower electrodes of the Penning trap, also not shown in the figures.  It boosts the axial magnetic field in the trap section where \pbar are initially loaded from 1 Tesla to 3.7 Tesla while antiprotons are being loaded, greatly increasing the \pbar loading efficiency.  The larger field also allows the electrons, used to cool the \pbar, to lose energy via synchrotron radiation with a time constant of 0.4 s rather than 4 s. This loading coil is turned off during the time that antihydrogen atoms are produced and trapped.

The Ioffe windings are very different from those in the first generation trap.  The use of many fewer windings gives the Ioffe coils a much lower inductance, as needed to inject and remove current much more rapidly. One consequence is that much higher current (up to 680 A) must be provided to achieve the same trap depth for \Hbar. Another consequence is that active quench detection and rapid external dissipation of the energy stored in the coils is required in order to protect the coils, as well as to allow removing the currents in tens of milliseconds. Another difference is that two sets of racetrack coils make it possible to apply either quadrupole or octupole Ioffe fields. In addition, bucking coils are added to make it possible to modify the pinch coil field -- to make the magnetic field vary less near the center of the trap, for example.

\begin{figure}
	\includegraphics*[width=\columnwidth]{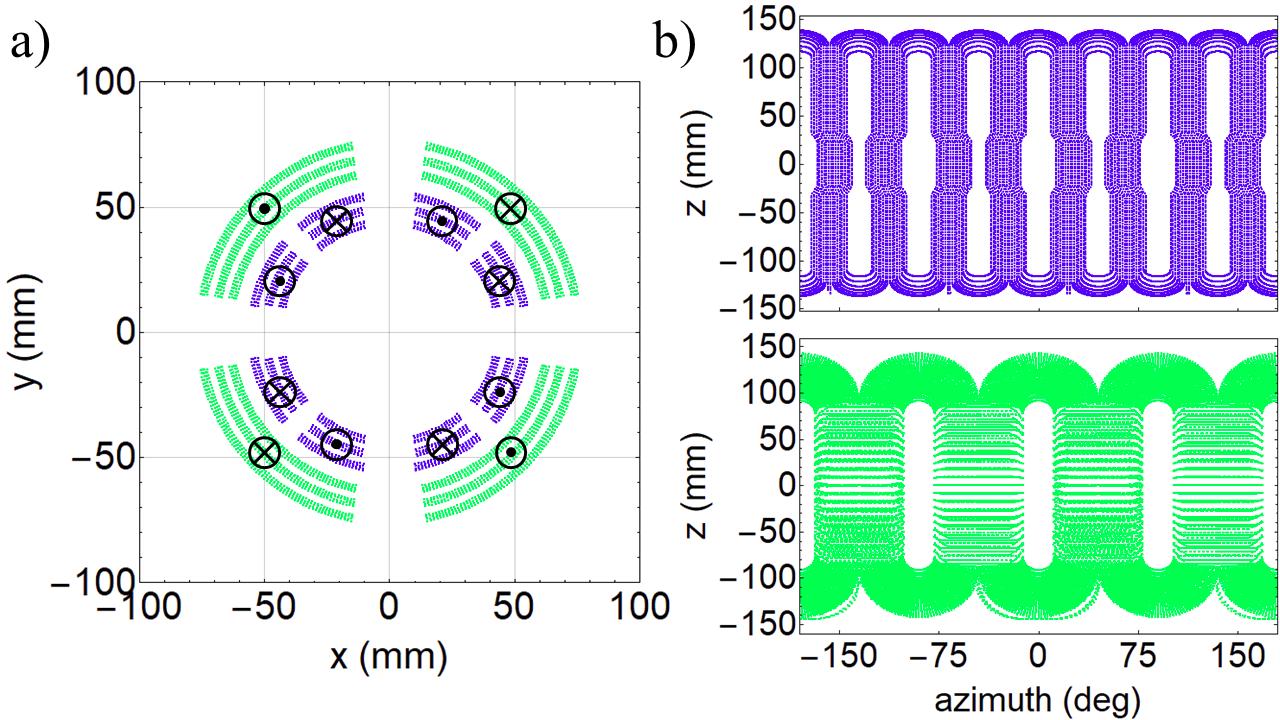}
	\caption{a) Cross section of the coil winding patterns for the racetrack coils.  The green outer set is the quadrupole coil, the blue inner set is the octupole coil. b) Rolled-out views of the octupole (top) and quadrupole (bottom) winding patterns. }  
	\label{fig:IoffeTrapWindings}
\end{figure}

The winding patterns for the new racetrack coils are shown in  Fig.~\ref{fig:IoffeTrapWindings}.  Fig.~\ref{fig:IoffeTrapWindings}a is a cross section of the winding patterns in the midplane. The $\otimes$ and $\odot$ signs indicate the relative current directions.  Fig.~\ref{fig:IoffeTrapWindings}b shows an unwrapped view of the windings. 

The quadrupole windings are outside the octupole windings because a quadrupole field falls off less rapidly with radius than does an octupole.  They include 744 ``vertical current bars.''  These are implemented as 4 layers of 56 superconducting wires, 4 layers of 62 wires and 4 layers of 68 wires. The vertical current bars are connected by the rounded contours shown in the figure.  The 4 sideports fit between the quadrupole windings.

The octupole windings, inside the diameter of the quadrupole windings, include 216 ``vertical current bars,'' implemented as 4 layers of 16 wires, 4 layers of 18 wires, and 4 layers of 20 wires. Near the midplane of the trap the wires are spread in four locations to make room for the sideports.    

The pinch and bucking coils are at a radius larger than that of the octupole and quadrupole.  So that they dominate the axial field, they are  closer to the center of the trap than are the connections between the vertical current bars.  The pinch coil has 988 turns, with 13 layers of 38 turns in its upper section being duplicated in its lower section.  The two sections are joined such that the current goes in the same azimuthal direction in each section.  Similarly, the bucking coil has 364 turns, with upper and lower sections each of which has 13 layers of 14 turns.  The current in each section goes in the same azimuthal direction.   

The advantage of the lower inductance is that it makes it possible to turn off the neutral particle traps a hundred times more rapidly -- essentially between the dark counts produced by cosmic rays that trigger the annihilation detectors used to tell us that a \pbar has annihilated when an \Hbar atom is released from the trap.  This trap turns off rapidly enough that most \Hbar will be released within 25 ms.  This is a big improvement on the 10 minute design value for the first generation trap, and on the 1 second realized by deliberately quenching the first generation trap. 

A related advantage is that the lower inductance makes it possible to bring Ioffe coils to full current in a time reduced 10-fold compared to that required by the first generation.  This was intended to speed the study of trapped antihydrogen generally. It was also intended to reduce the losses of trapped antiprotons and positrons that occur before antihydrogen forms, and hence increase the amount of trapped antihydrogen.   

Since the wires are insulated from each other using formvar varnish, the inductance values (see Table~\ref{table:Measured}) can be measured at room temperature to several percent using a standard inductance bridge using a low frequency AC signal (1 kHz).  At room temperature the current flows through the copper of the wire, while at 4 K the current flows through the superconducting strands within the copper.  This variation within the small wires should not greatly affect the inductance that is measured.  

A low inductance nested Penning-Ioffe trap for antihydrogen atoms poses significant technological challenges.  Many hundreds of amperes of current not only flow in the superconducting wires of the Ioffe coils, but these large currents are turned on and off very rapidly.  Great mechanical stability is required because of the large and impulsive forces between the windings.  The 4 K coils are only 6 mm away from Penning trap electrodes cooled to 1.4 K. The energy stored in the inductance of the superconducting coils is much greater than that needed to destroy the coils if this energy is turned into heat in  a quench.  

\begin{figure}
	\centering
    \includegraphics[width=\columnwidth]{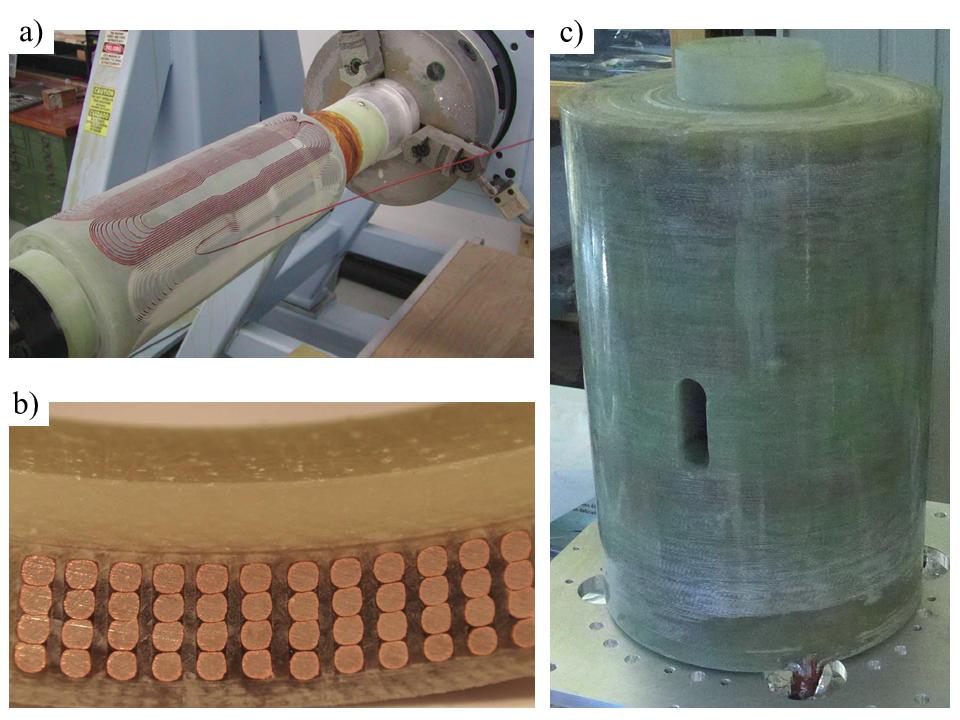}
    \caption{The Ioffe windings were set into grooves machined into a G10 epoxy fiberglass tube (a) and then epoxied into those grooves (b).  The end result is a solid epoxy-fiberglass structure with embedded Ioffe coils, featuring an open tube in its central axis for Penning trap electrodes, 4 radial slots for sideports, and small channels between coil layers to improve LHe cooling (c).}
	\label{fig:CoilWiresPictures}
\end{figure}

To achieve the required mechanical stability, a direct-wind construction method was employed in the fabrication by Advanced Magnet Lab (AML). Fig.~\ref{fig:CoilWiresPictures} shows how the racetracks were fabricated.  The Ioffe windings were set by hand into grooves machined into a G-10 epoxy fiberglass tube (a) and the windings were then epoxied into the grooves (b).  A new layer of epoxy fiberglass was then fabricated over these windings, grooves were cut into this new layer, and more wires were deposited and epoxied.

The octupole and quadrupole windings are made with round superconducting wire with a 0.85 mm diameter and a 0.9:1 cross-sectional ratio of copper to NbTi (Supercon 56S53). The pinch and bucking coils are wound using rectangular wire that is 1.576 mm by 1.052 mm, including formvar varnish insulation that is 0.035 mm thick.  The cross-sectional ratio of copper to superconductor in this wire is 2.7 to 1. 


The inductance of each half of the coil, with respect to a roughly centered tap attached to the coil, is similarly measured.  The mutual inductance between halves of each coil is determined by the measurements of $L$, $L_1$ and $L_2$ using
\begin{equation}
M=\frac{1}{2}(L-L_1-L_2).
\end{equation}
The series resistance is only useful as a room temperature diagnostic, since it goes to zero when the wires become superconducting at 4 K.

\begin{table}
	\caption{Inductances in H and resistance in ohms, measured at 300 K.}
	\begin{tabular}{rcccccc}
		& ~~$L$~~ & ~~$L_1$~~  & ~~$L_2$~~  & ~~M~ & $R_s(300 K)$   \\ 
		\hline  \noalign{\vskip 2mm} 
		octupole    & 0.019 & 0.0060 & 0.0065 & 0.0032 & 31    \\ 
		quadrupole  & 0.113 & 0.037 & 0.030  & 0.0229 & 53    \\ 
		pinch       & 0.107 & 0.052 & 0.051 & 0.0021 & 9.4  \\ 
		bucking     & 0.020 & 0.0087 & 0.0087 & 0.0012 & 3.4  
	\end{tabular}
	\label{table:Measured}
\end{table}

The superconducting Ioffe coils are connected vertically up to room temperature current leads though conducting sections. The superconducting wires of each coil are soldered to bus bars, 62 cm long, that are thick copper conductors between which high temperature superconducting tape (American Superconductor BSCCO) is sandwiched.  

As long as any length of the bus bar remains immersed in 4 K liquid helium, the copper part of the bars will keep the superconducting tape below its critical temperature.  This tape can typically carry 110 A of current at 77 K, in the absence of a background magnetic field.  They are mounted in orientations that minimize the component of the background 1 T field perpendicular to the tape, which would otherwise limit the critical current.  The other end of each bus bar is connected to a vapor cooled lead (VCL) that extends out of the dewar to a room temperature connection.  Cold helium vapor is directed through channels in each VCL to keep heating from increasing both their resistance and their power dissipation for a given current.  The bus bars and VCLs are custom assemblies manufactured by Cryomagnetics Inc.

\subsection{Equipotentials for Ground State \Hbar}
\label{sec:HbarPotentials}

The equipotentials of Ioffe traps are surfaces of constant U, which is equivalently a constant magnitude for the vector sum of the spatially uniform bias magnetic field and the magnetic field from the Ioffe windings (see Eq.~\ref{eq:PenningIoffeField}). \Hbar and H atoms have a magnetic moment of 1 Bohr magneton, $\mu_B$, and the corresponding potential energy of low-field-seeking atoms is $-\mu_B|\vec{\bf B}|$ with respect to the magnetic minimum.  Because the energies are so small, we divide by the Boltzmann constant ($k$) to represent them in mK temperature units.  The conversion for the trap depth from Tesla to mK is $\mu_B/k = 672$ mK/Tesla.

In what follows, the equipotentials for the deepest symmetric octupole and quadrupole traps that have been achieved so far are illustrated. By symmetric trap we mean that we have chosen to make the axial well depth equal to the radial well depth (which is determined by the field strength at the inner surface of the trap electrodes).  The currents for these traps are listed in Table~\ref{table:DeepestTrap}.  The Penning-Ioffe traps include a 1 Tesla bias field directed in the $-\hat{\bf z}$ direction.  These maximum currents, established by repeated trials with higher and higher currents, are the highest for which robust operation was achieved without quenching.  

\begin{table}
	\caption{Ioffe coil currents in A for the deepest symmetric Penning-Ioffe traps realized in the presence of a 1 Tesla bias field.}
	\resizebox{\columnwidth}{!}{
	\begin{tabular}{rcccc}
	&  octupole & quadrupole & pinch & bucking \\ 
		& coil & coil & coil & coil\\
		\hline \noalign{\vskip 2mm}  
		octupole trap    &  680 & 0 & 210 & -179 \\ 
		quadrupole trap  &  0 & 470 & 310 & -264\\ 
	\end{tabular}}
	\label{table:DeepestTrap}
\end{table} 

\begin{figure}
	\centering
	\includegraphics*[width=\columnwidth]{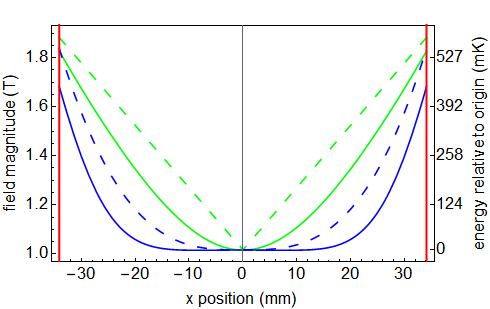}
	\caption{The field magnitude and potential energy at $z=0$ for an \Hbar atom near the center of the ATRAP quadrupole (green) and octupole (blue) Penning-Ioffe traps with a 1 Tesla bias field. Dashed curves are $|x|$ and $|x|^3$ dependencies for comparison. The red lines are at the inner radius of the Penning trap electrodes.}  
	\label{fig:RadialWells}
\end{figure}

\begin{figure}
	\centering
	\includegraphics*[width=\columnwidth]{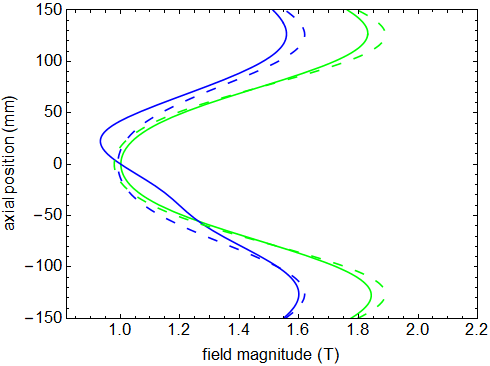}
	\caption{The field magnitude on the z axis for quadrupole (green) and octupole (blue) Penning-Ioffe traps with a 1 Tesla bias field.  The dashed lines are calculations for idealized versions of the windings (without sideports or connections between current bars).}  
	\label{fig:AxialWells}
\end{figure}

The field profile in the midplane of the Penning-Ioffe trap, perpendicular to its axis, is shown in Fig.~\ref{fig:RadialWells}.  The axial well for \Hbar atoms along the z-axis is shown in Fig.~\ref{fig:AxialWells}.  Because of the vector addition of the bias field and the field from the pinch and bucking coils, the potential energies of the Penning-Ioffe traps do not manifest the $\rho$ or $\rho^3$ shape discussed earlier (the dashed curves in the figure). They vary instead approximately as $[1+ (B_R/B_0)^2\, (\rho/R)^2]^{1/2}$ or $[1+ (B_R/B_0)^2\, (\rho/R)^6]^{1/2}$ under the assumption that the net axial field at the center from the pinch and bucking coils is very small.

\begin{figure}
	\centering
	\includegraphics*[width=0.48\columnwidth]{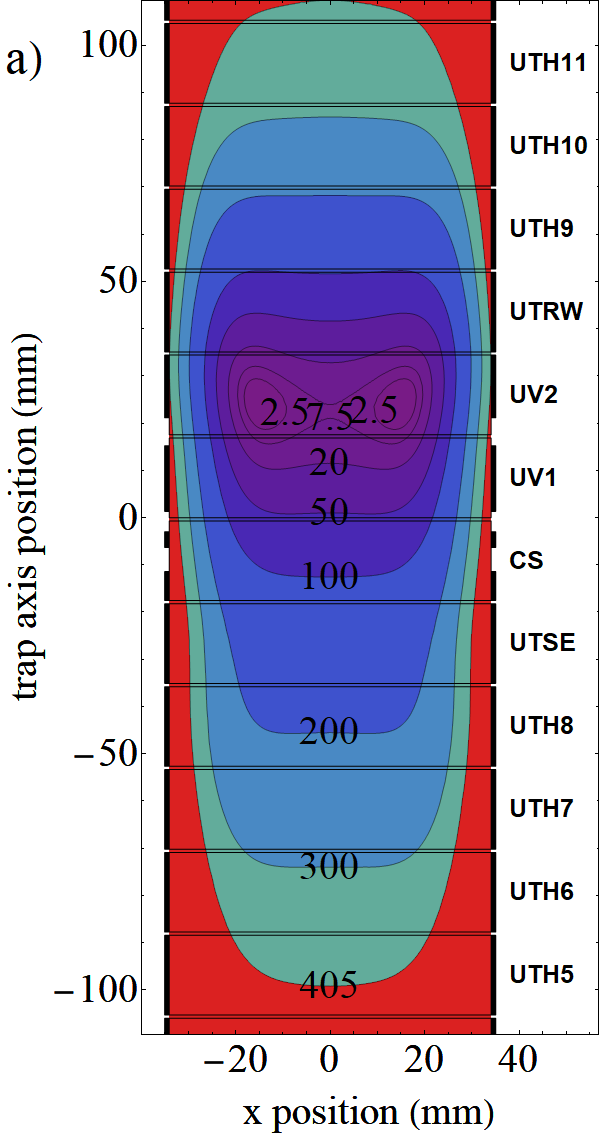}\hfill
	\includegraphics*[width=0.48\columnwidth]{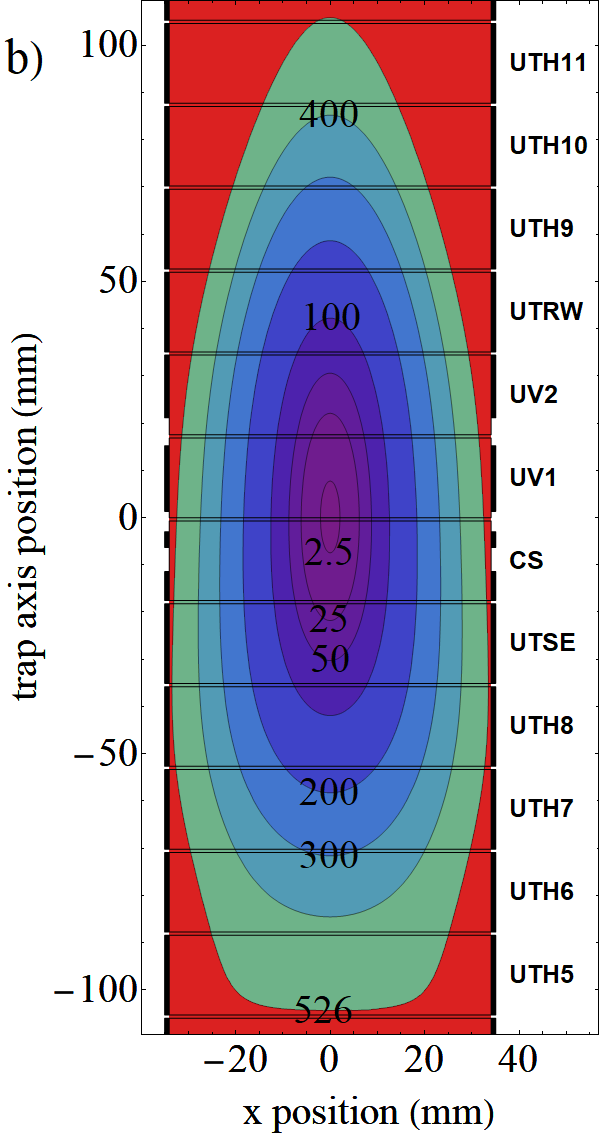}
	\caption{Energy contours in mK for the ATRAP octupole trap (680 A octupole, 210 A pinch, -179 A bucking) (left) and for the ATRAP quadrupole trap (470 A quadrupole, 310 A pinch, -264 A bucking) (right).  \Hbar atoms with energy below 2.5 mK laser-cooling limit would be within the lowest energy contour.  Labels along the right are the names we assigned to the individual Penning-trap electrodes.}       
	\label{fig:OctupoleAndQuadrupoleContours}
\end{figure}

Fig.~\ref{fig:OctupoleAndQuadrupoleContours} shows the equipotential energy contours for the symmetric quadrupole and octupole traps. A notable feature of the equipotentials is the lack of symmetry above and below the center of the trap along the trap axis.  This may at first seem surprising given the symmetry of the coils. This asymmetry arises primarily because the bias field that points in the $-\hat{\bf z}$ direction is added to the field produced by the current-bar connections at the top and bottom of the trap.  The direction of the current in these connecting segments alternates from one bar to the next, so this part of the trap asymmetry reverses with the rotational symmetry of the coils.  For example, the field contours in the quadrupole trap in the yz-plane are the mirror image (reflected over z=0) of those in the xz-plane.

The vertical current bars for the octupole are slightly distorted near the center of the trap to accommodate the four side windows.  The result (notable in Figs.~\ref{fig:AxialWells} and \ref{fig:OctupoleAndQuadrupoleContours}a) is an axial shift in the location of the shallow 7.5 mK deep toroid that contains the circle of minimum magnetic field.  The toroid field minimum for the octupole trap with side windows has its minimum magnetic field about 15 mm off axis at a height that differs by 3 mm from that of the minimum field on the central axis, which itself is displaced from the geometric center of the Ioffe trap by about 22 mm.  A set of laser access holes in the Penning trap electrodes are positioned to allow a laser to address \Hbar located in the toroidal trap minimum. 

When antihydrogen is first trapped, its energy in temperature units should be distributed between 0 and the 400 to 500 mK depth of the traps.  Before laser cooling, the stored antihydrogen atoms should be distributed within the much larger outer contours in Fig.~\ref{fig:OctupoleAndQuadrupoleContours}. 

Of particular interest are the 2.5 mK contours near the center of the trap.  Laser cooling via the 121 nm transition between the 1s and 2p levels of antihydrogen and hydrogen has a cooling limit of about 2.5 mK. After optimal laser cooling, the antihydrogen atoms stored in the trap should be mostly within the 2.5 mK contour.  For the octupole case, this inner contour is a torus as a result of the relative magnitudes of the radial components of the octupole and pinch coils.  The shape and position of this 2.5 mK ``equipotential'' can be modified somewhat by increasing the field from the pinch and/or bucking coils.

\subsection{Vacuum Enclosure}

\begin{figure}
    \centering
    \includegraphics*[width=\columnwidth]{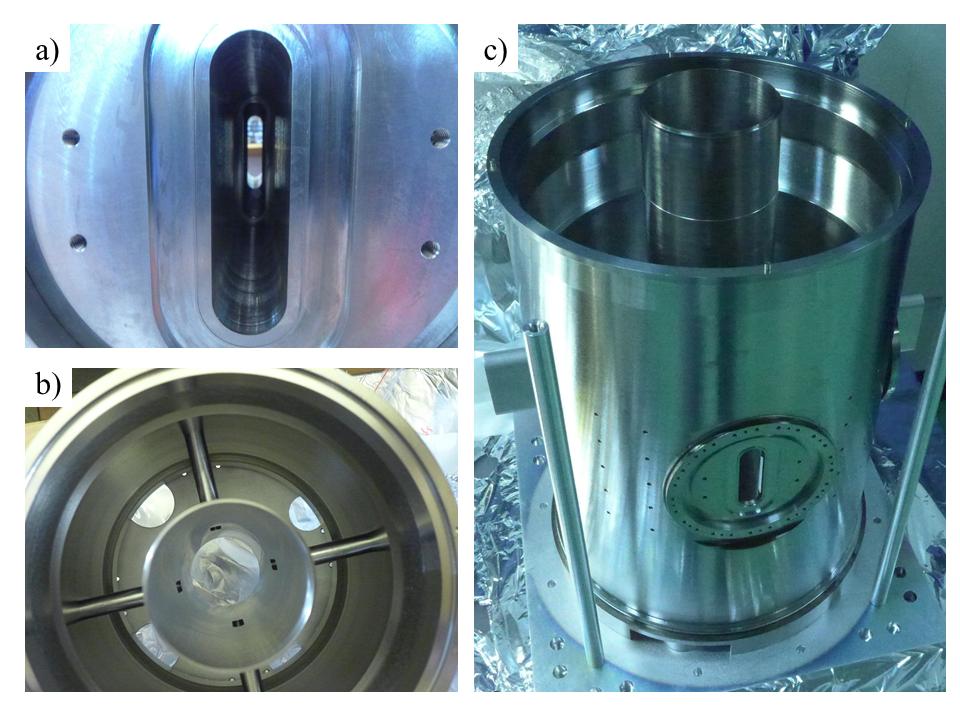}
    \caption{(a) Sideport oval with view through the Ioffe trap.  The oval access to the center is 8.9 mm wide and 45.7 mm high. (b) Test assembly of the enclosure components showing the sideport tubes connecting the outer cylinder to the inner cylinder. (c) The vacuum enclosure partway through the welding process.  An aluminum jig inserted through the siderport holes ensures the inner and outer cylinders are properly aligned during e-beam welding.}
    \label{fig:VacuumEnclosure}
\end{figure}

In addition to their effect on the octupole trapping potential, the four sideports (Fig.~\ref{fig:IoffeTrapCrossSection}) also add to the challenge of constructing a vacuum enclosure (Fig.~\ref{fig:VacuumEnclosure}) around the G10 winding assembly.  The enclosure separates three different volumes: the exterior insulating vacuum, the trapping volume in the central bore, and the volume to be filled with liquid helium to keep the NbTi wires of the Ioffe coils at 4 K.  Advanced Magnet Lab (AML) manufactured the G10 winding assembly, and ATRAP took on the task of constructing a vacuum enclosure around it.

\begin{figure}
	\centering
	\includegraphics*[angle=90,height=0.6\columnwidth]{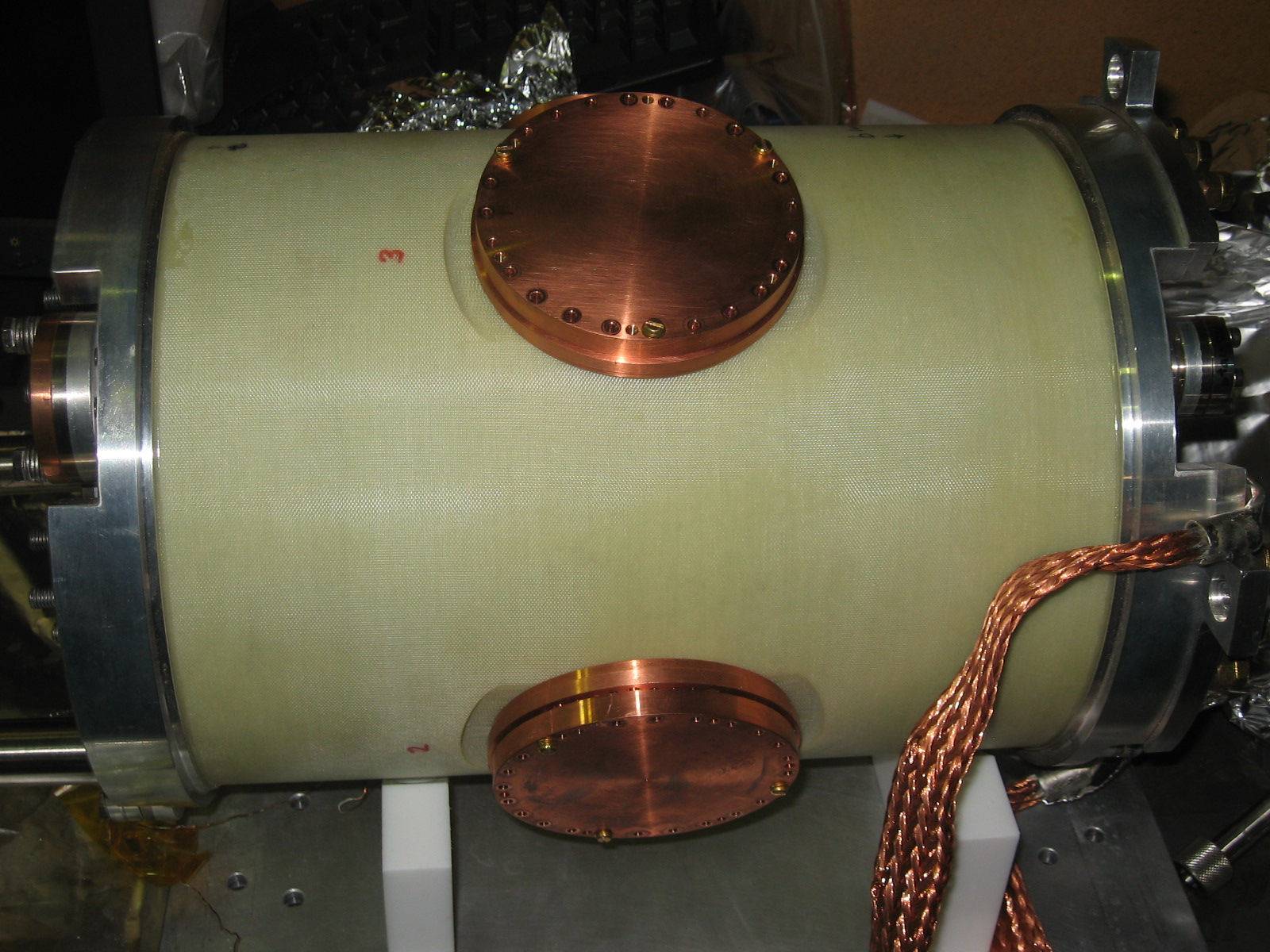}
	\includegraphics*[height=0.6\columnwidth]{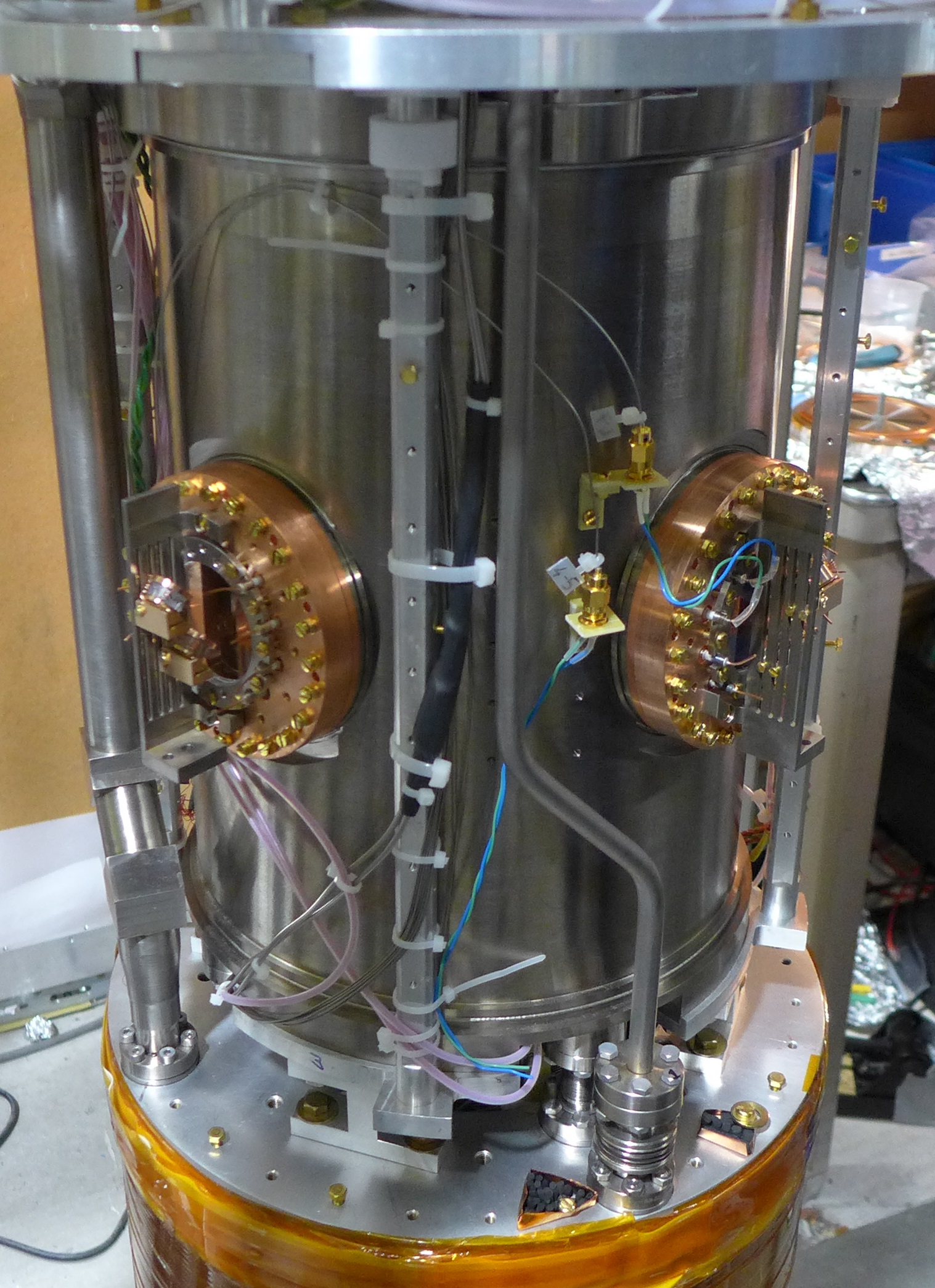}
	\caption{A G10 epoxy fiberglass enclosure (left) and a Ti enclosure (right), both designed to contain the liquid helium that cools the Ioffe trap coils within the enclosures. The picture of the titanium enclosure also features sideport flanges with windows and mirrors for sending laser light through the trap radially.}  
	\label{fig:G10AndTi}
\end{figure}

The first attempt (pictured on the left of Fig.~\ref{fig:G10AndTi}) was a mixed-material vacuum enclosure: top and bottom aluminum plates with welded-in bi-metal CF flanges (Atlas Technologies), epoxied to G10 components comprising the outer cylinder and sideport tubes.  Copper rings were epoxied outside the sideports for use in forming the indium seals for the flanges that would separate the trap vacuum from the insulating vacuum.  One advantage of using an insulator for the central parts of the vacuum enclosure was that the rapid changes in the Ioffe trap magnetic fields could not produce the substantial eddy currents that would flow in a metal system.  Eddy currents are difficult to calculate, and they can modify both the time structure and the spatial distribution of the fields that the coils produce.  Another advantage was that the Penning trap electrodes could be mounted against the G10 of the Ioffe trap bore without risk of shorting to ground as would happen if the electrodes contacted a metal enclosure.

\begin{figure}
	\centering
    \includegraphics*[width=\columnwidth]{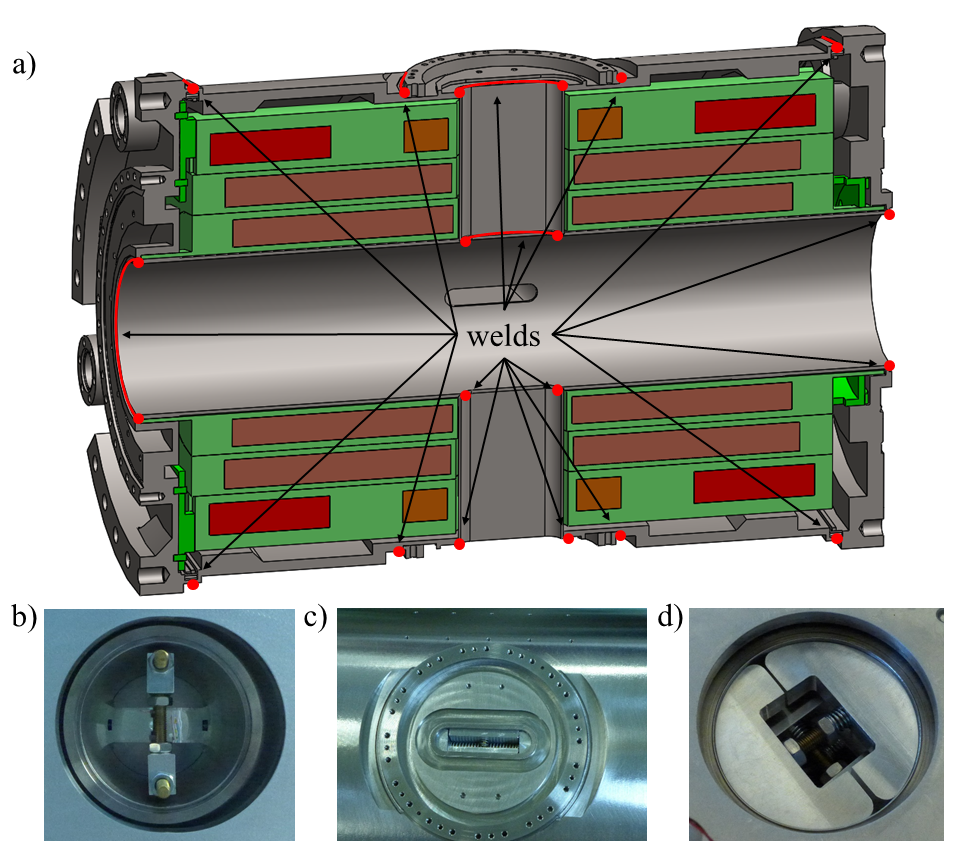}
	\caption{A diagram of the e-beam weld locations.  Red dots indicate where the cross section of the model cuts through the side of an e-beam weld, and red curves follow the weld path where visible from this angle (a).  The photos show several of the heat sinks used to avoid heat damage to the G10 encapsulating the coils, located at the interior of the sideports (b), the exterior of the sideports (c), and the inner bore by an end plate (d).}  
	\label{fig:eBeamWelds}
\end{figure}

After several attempts at cooling down the apparatus with the G10 vacuum enclosure installed resulted in cracked epoxy joints and thus leaks from the liquid helium space into both the insulating vacuum and the trap vacuum, we embarked on a redesign of the enclosure using metal components welded around the G10 block of coil windings. Stainless steel was considered for this enclosure as its resistivity remains high even at cryogenic temperatures, damping the eddy currents produced by rapid changes in the Ioffe trap's magnetic field.  Because the apparatus was designed for extremely precise antihydrogen spectroscopy, however, we decided against stainless steel.  Even the nominally non-magnetic alloys of stainless steel have some residual magnetism, and more can develop over time due to cyclic stresses of the sort expected to accompany thermal cycling between 300 K and 4 K.

In order to avoid the drawbacks of stainless steel, the possibility of using titanium (Ti) alloys was investigated, specifically those listed in Table \ref{table:TiAlloys}.  Ti alloys are stronger than pure Ti, making it possible to use less material between the Ioffe coils and the interior volume of the Penning traps.  The additional material between the winding assembly bore and the Penning trap electrodes requires the electrode radius (and thus the magnetic trap depth) to be reduced accordingly, so a thinner wall for the bore of the vacuum enclosure means a deeper \Hbar trap.

Alloys (unlike pure Ti) also maintain a substantial resistivity as they cool to cryogenic temperatures. Material research for cryogenic use led to the development of the extra-low interstitial (ELI) varieties of grades 5 and 6.  Interstitial elements (impurities such as oxygen, nitrogen, and carbon) strongly affect the tensile properties of titanium alloys at low temperatures, and can make them prone to brittle fracture.  While it is not commonly used in cryogenic systems, the  impurity levels for grade 9 titanium are below the limits of the grade 5 ELI specifications;  it should be equally unlikely to become brittle at our operating temperature.

\begin{table}
	\caption{Candidate Ti alloys for the Ioffe trap enclosure}
	\begin{tabular}{c| c c c}
		\hline
		Ti alloy&~~~~~Al~~~&~~~V~~~&~~~Sn~~~\\
		\hline
		Grade 5 & 6\% & 4\% & 0\% \\ 
		Grade 6 & 5\% & 0\% & 2.5\% \\ 
		Grade 9 & 3\% & 2.5\% & 0\% \\ 
	\end{tabular} 
	\label{table:TiAlloys}
\end{table}

A superconducting enclosure could unacceptably distort the magnetic field of the Ioffe trap in a way that would be difficult to precisely calculate or measure. With a concern that some of the titanium alloys, due to their vanadium content, would become superconducting at 4 K, those under consideration were tested. The tests for superconductivity were performed using a SQUID magnetometer (Quantum Design MPMS XL) at Harvard's Laukien-Purcell Instrumentation Center.  Three different pieces of grade 5 titanium and one each of grade 6 and grade 9 were obtained and tested. 

For each sample, a magnetic field generated by the device was incremented from zero to 15000 Oe (in vacuum, B = 1 G corresponds to H = 1 Oe), then to 15000 Oe in the opposite direction, and finally back to zero. At each field value, the sample's magnetic moment was measured. As the sample had a known geometry, the magnetic moment per unit volume (magnetization) could be deduced from this measurement. The magnetization was measured at 3.5 K and 4.2 K.

At 4.2 K, all three grade 5 samples generated the familiar magnetization curve of a type-II superconductor. The grade 6 and grade 9 samples, on the other hand, appeared simply to be paramagnetic.  At 3.5 K, the grade 6 sample still showed no sign of superconductivity. However, a small, difficult-to-resolve feature appeared near zero field in the magnetization curve for the grade 9 sample. This feature was repeatable, and a careful analysis of background sources could not account for it. Nevertheless, for the fields and temperatures our enclosure would experience, the grade 9 sample did not appear to behave like a superconductor.

Grade 6 titanium outperformed the other two alloys in the magnetometer tests, so it was the obvious choice. However, due to difficulty sourcing an appropriate amount of grade 6 with material certifications, grade 9 was chosen instead. Two samples from each of the five pieces of grade 9 titanium stock obtained were tested. Samples from the same piece of titanium had similar magnetization curves, but there were significant differences between samples from different pieces. In particular, four of the five samples behaved as superconductors at 4.2 K, although critical fields and critical temperatures (measured at 25 Oe) varied between the four. Even so, the 4.2 K critical field for each piece was comfortably less than the minimum field that piece would see during normal operation.

With the raw materials in hand and understood, the Harvard shop machined them according to the designs we provided.  The finished components and the G10 coil winding assembly were then brought to Joining Technologies, who used electron-beam welding (performed in a vacuum chamber) to join the pieces of the titanium enclosure together. Fig.~\ref{fig:eBeamWelds} shows the locations of the e-beam welds.  Temporary clamps, also pictured in the photos in the figure, positioned the pieces for welding.  Additional aluminum heat sinks were included to prevent heat damage to the G10 block encasing the Ioffe coils.  Avoiding heat damage during welding was challenging given that a precise alignment of the Ioffe coils with the enclosure required contact between the titanium and G10. It was also crucial that differential thermal contraction between Ti and G10 during cooling to 4 K would not break the G10 in contact with the Ti or crack the Ti weld joints.  If the radial access ports had not been included, constructing the metal vacuum enclosure would have been much easier.  Additional challenges due to the sideports included implementing a precise rotational register to ensure the G10 sideport holes align with the titanium sideport tubes, and performing the internal welds connecting the 4 sideports to the central tube since the electron beam had to be directed into the small sideports from the outside.

Due to the layer of titanium in the central bore, the octupole trap depth is 20\% lower than would have been possible with a G10 enclosure, and the quadrupole trap is 10\% lower.  To date the grade 9 Ti vacuum enclosure (pictured on the right of Fig.~\ref{fig:G10AndTi}), as part of ATRAP's antihydrogen apparatus, has been successfully cycled between 300 K and 4 K eight times and has spent 67 weeks total at 4 K.



\subsection{Electrical Circuit}

\begin{figure}
	\centering
	\includegraphics*[width=\columnwidth]{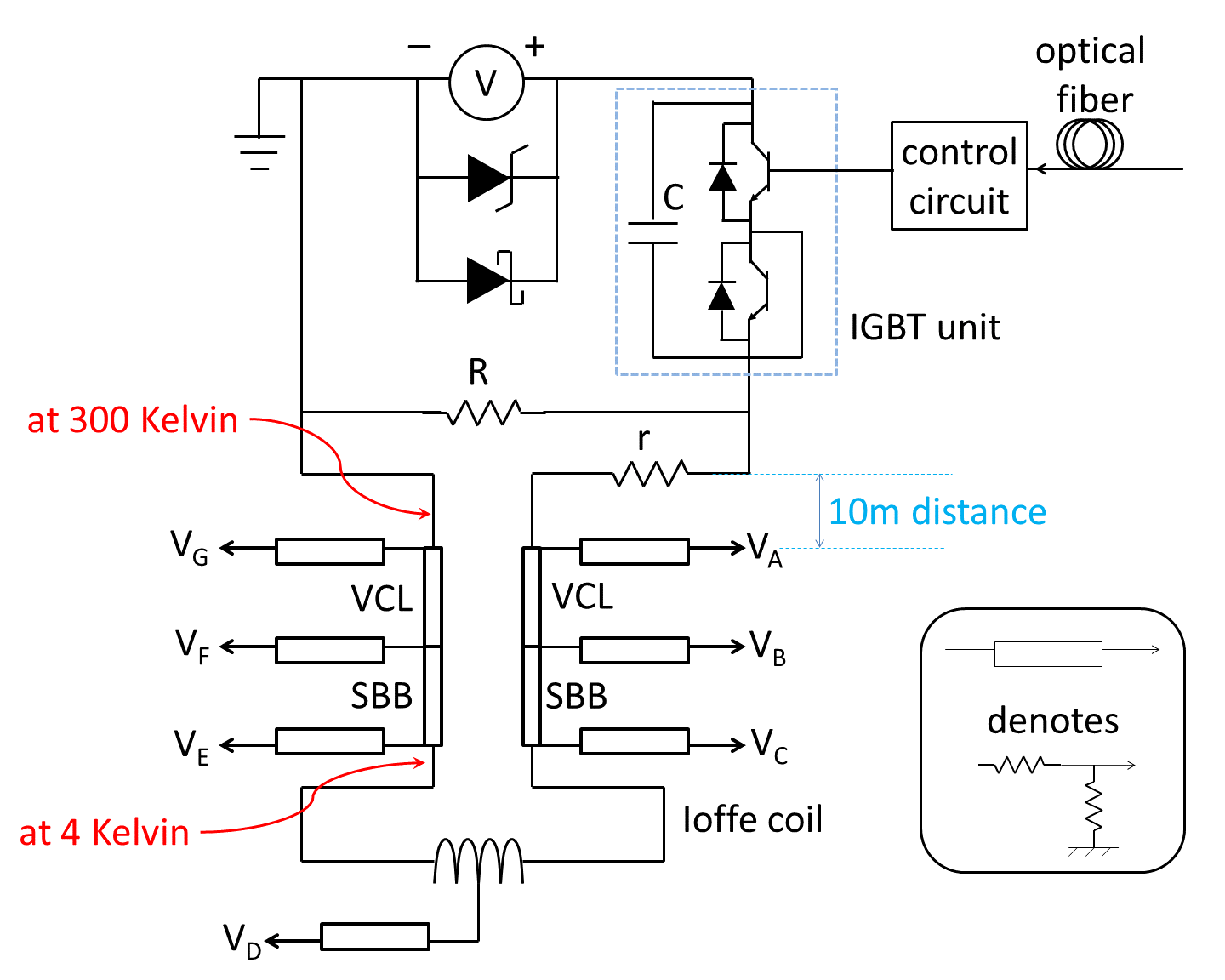}
	\caption{Schematic of a coil electrical circuit, consisting of a room-temperature power supply, IGBT and dump resistor connect to a 4 K Ioffe coil through vapor-cooled leads (VCL) and superconducting bus bars (SBB).  Tap voltages ($V_A$, ..., $V_G$) are reduced and monitored at room temperature. }  
	\label{fig:FullCircuit}
\end{figure}

Each coil in the Ioffe magnet is part of a larger circuit designed to meet the requirements of our antihydrogen program: fast ramp-ups to the hundreds of amps needed to fully energize the trap, fast removal of current on-demand to release trapped \Hbar within a narrow time window, and fast removal of current in the case of a detected quench of the coils.  Fig.~\ref{fig:FullCircuit} shows a schematic including all of the components of a coil's circuit.  There is one such circuit for each coil, with the power supply configuration and the value of the parallel resistance differing depending on the properties of the particular coil.

The power supplies used are HP model 6681A, which in the standard configuration can output 8 V and 580 A.  The pinch and bucking coil circuits each use one of these supplies, while the quadrupole coil circuit uses a supply configured for 7 V and 650 A.  Two 6681A supplies are wired in parallel to provide the current for operating the octupole coil.  To protect these devices against both forward and reverse voltages, transient voltage suppression diodes and Schottky diodes are placed in parallel with the power supplies.

When the IGBT (insulated-gate bipolar transistor) unit is enabled, current flows from the positive terminal of the power supply through the upper transistor in Fig.~\ref{fig:FullCircuit}.  This unit (Semikron SKiiP 1513GB172-3DL) is integral to the capability to remove the current quickly, being both fast-switching ($\mu$s scale) and rated for both high currents and high voltages.  The high-current capability is necessary for steady-state operation.  When the switch is opened to dump the current, the IGBT unit must be able to survive the resulting inductive voltage spike as well.  A capacitor in parallel with the transistors prevents charge pileup and potentially damaging voltage spikes immediately following the opening of the IGBT.  While the IGBT units themselves are rated for 1700 V, the capacitors are only rated for 1100 V.  

\begin{figure}
	\centering
	\includegraphics*[width=\columnwidth]{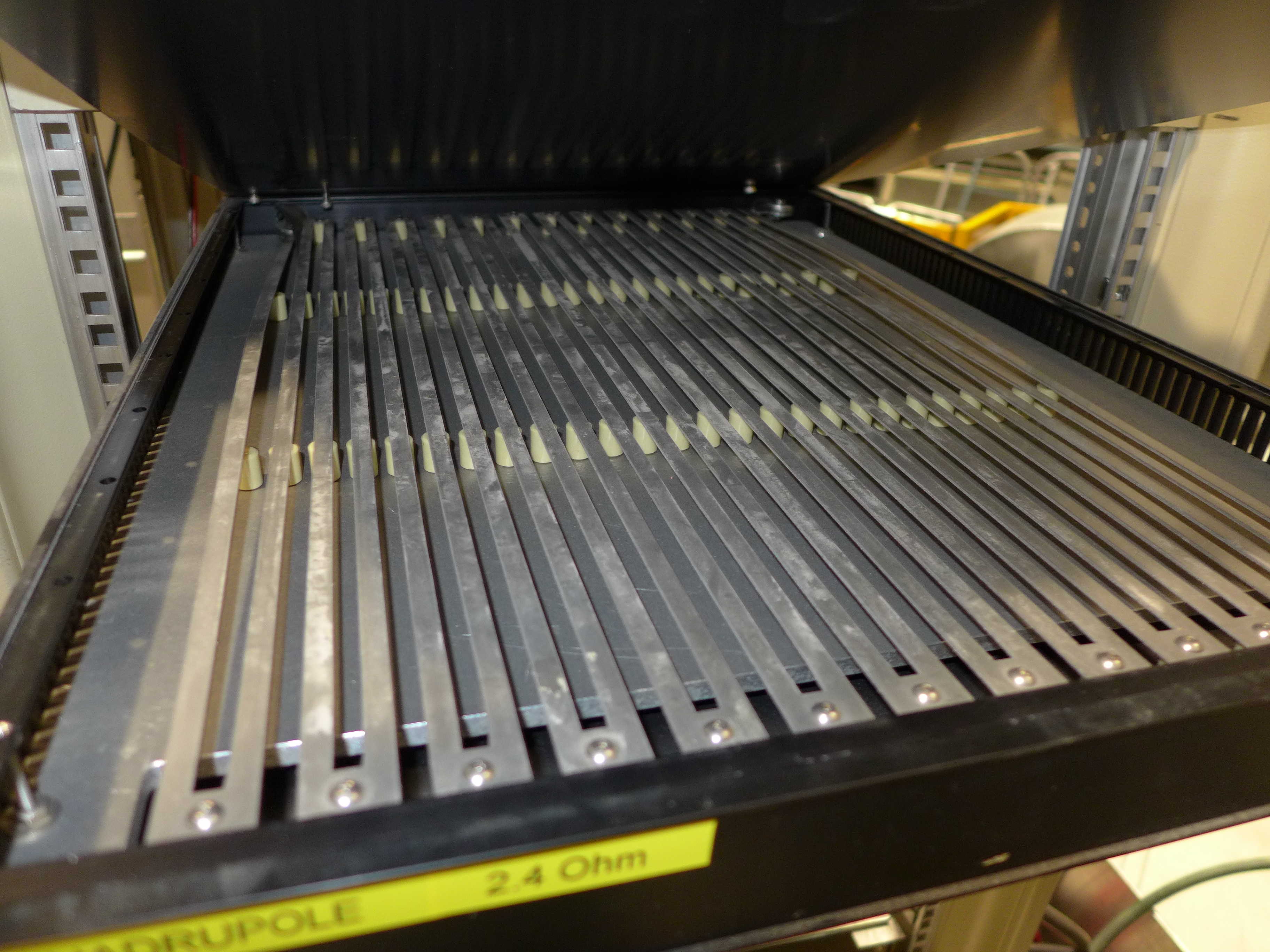}
		\caption{Resistor constructed from layers of thin sheets of stainless steel.}  
	\label{fig:ResistorStack}
\end{figure}

A dump resistor (labelled R in Fig.~\ref{fig:FullCircuit}) wired in parallel with the magnet coil dissipates the energy stored in the coil when the IGBT opens.  These dump resistors were constructed by machining 1/8"-thick stainless steel sheets to make long path lengths from one terminal to the other, then stacking the number of these sheets that results in the desired resistance (see Fig.~\ref{fig:ResistorStack}).  The voltage spike generated when the current is removed from the coil is a function of the coil inductance, the initial coil current, and the resistance of the dump resistor.  For each coil, a dump resistor was built such that the peak voltage when the coil starts at maximum operating current will be safely less than the 1100 V rating of the IGBT unit's capacitor.  The measured values of $R$ and $C$ are listed in Table~\ref{table:MeasuredRC}.

To carry high currents from the output of the IGBT unit to the antihydrogen apparatus and back to the negative terminal of the power supply, 10 meter long copper wires of size 700 kcmil (about 350 $\textrm{mm}^2$ cross section) are used.  These wires have a current-carrying capacity of 520 amperes, sufficient for the quadrupole, pinch, and bucking coils' maximum currents.  Since the maximum operating current of the octupole coil is 680 amperes, two sets of these wires are used for the octupole, one from each of the parallel power supplies.

To allow for more-flexible cables to make the connection to the antihydrogen apparatus, the 700 kcmil wires end at a junction box near the apparatus.  There, they connect to flexible copper-braid straps (Storm Power Components' FlexBraid product series) with lengths ranging from 1.5 meters to 2.7 meters.  The pair of wires from the octuople supply connect to a strap of 1.5 by 3.5 cm cross section, rated for 900 Amps.  The straps used for the quadrupole are rated for 600 Amps (1.0 by 3.5 cm cross section), and those for the pinch and bucking coils are rated for 470 Amps (0.7 by 3.0 cm cross section).

These high-current straps are bolted to copper flags at the top of the antihydrogen apparatus.  These flags are in turn clamped to the ends of the appropriate vapor-cooled lead (VCL) which carries the current down into the cryogenic space where it passes through HTS busbars and on into the superconducting magnet coils, as described in detail in Section \ref{sec:IoffeCoilSection}.

\subsection{External Protection Circuit}
\label{sec:qpCircuit}

In order to drain the current from a coil in the event of a quench, we need to be able to detect that a quench has begun.  A quench occurs when a small part of the superconducting wire transitions to the normal-conducting state.  The high current running through the non-zero resistance of this section of wire causes heating, which results in the surrounding wire becoming normal as well.  Thus a chain reaction occurs, with our system having the potential to reach temperatures that would damage the magnet.  The appearance of a small section of resistive wire is accompanied by the appearance of a voltage drop across that section of wire, which is what is used to diagnose the beginning of a quench and trigger the subsequent removal of the current before it can cause damage.

To monitor voltage drops across several sections of the cryogenic part of the coil circuit, wires are soldered to each of the junctions to act as voltage taps.  As shown in Fig.~\ref{fig:FullCircuit}, one is located at the top of each VCL (A and G), one where each VCL connects to a busbar (B and F), and one where each busbar connects to the magnet coil (C and E).  The magnet manufacturer (AML) provided a voltage tap for each coil, connected to the center of the winding and referred to as tap D.  Our quench detection system is designed to be most sensitive to a difference in the voltage drop between each half of a magnet coil, which would arise if a section of one half of that magnet coil became normal-conducting.  This is preferable to looking for a change in the voltage drop over the entire magnet since it precludes the need to lower the sensitivity of the protection system while energizing a coil.  While charging the magnet induces a voltage over the full coil, it is symmetric between halves of the coil and thus would not trigger the protection system (unless a quench occurs while charging).

The voltage taps B through F are located in the cryogen space, so their signal must pass through an electrical feedthrough.  To eliminate the risk of high voltage on these taps causing an arc between pins of the feedthrough, the tap voltages first pass through a voltage divider that scales them down by a factor of five.  For symmetry, taps A and G are divided down externally before being passed to the quench protection electronics along with the rest of the voltage tap signals.

Our quench protection electronics continuously monitor the taps for voltage drops that exceed user-defined thresholds.  When that threshold is exceeded, these electronics send a signal through the optical fiber in Fig.~\ref{fig:FullCircuit} to open the IGBT, send a signal to the power supply to disable its output, and switch the state of a TTL output.  This system also has TTL inputs to allow the current to be dumped on command in \Hbar trials.  The TTL output level is monitored by a voltage tap datalogger and our particle detector electronics.  Higher-rate voltage tap data is recorded during a period before and after a fast current dump.  By including the TTL output state in the particle detector data, the times of detected events can be synchronized with the state of the neutral particle trap.

\begin{table}
	\caption{Resistance and capacitance values used in the coil electrical circuits.}
	\begin{tabular}{rcc}
		& ~~~~$R$ ($\Omega$)~~~~ & ~~~~$C$ (F)~~~~  \\ 
		\hline
		octupole     & 1.17 & 0.00159   \\ 
		quadrupole   & 1.99 & 0.00159   \\ 
		pinch        & 3.94 & 0.00164 \\ 
		bucking      & 1.90 & 0.00158 
	\end{tabular}
	\label{table:MeasuredRC}
\end{table}

\begin{figure}
	\centering
	\includegraphics*[width=0.5\columnwidth]{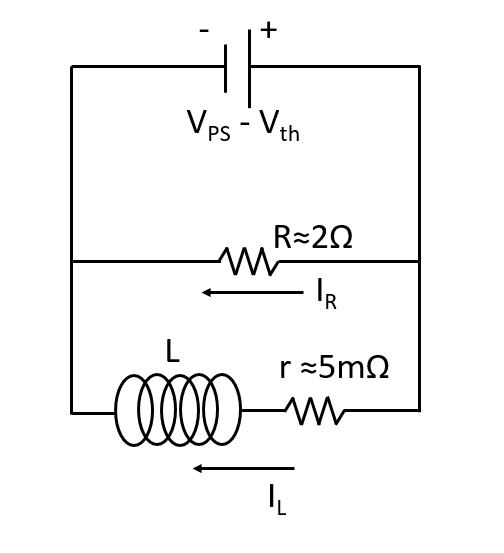}
	\caption{The effective circuit for charging each Ioffe coil.}  
	\label{fig:EffectiveCircuitCharging}
\end{figure}

\begin{table}
	\caption{Deduced time constants of the Ioffe circuits, as well as the ratio of the inductive voltage drop over the ``halves'' of each coil on either side of the central voltage tap $V_D$.}
	\begin{tabular}{rccc}
		& ~~~~$\tLR$ (s)~~~~ & ~~~~$\tRC$ (s)~~~~ & ~~~~$V_1/V_2$~~~~ \\ 
		\hline
		octupole    &  0.016 & 0.0019 & 0.95 \\ 
		quadrupole  &  0.057 & 0.0032 & 1.14 \\ 
		pinch       &   0.027 & 0.0065 & 1.01 \\ 
		bucking     &   0.010 & 0.0030 & 1.00
	\end{tabular}
	\label{table:Deduced}
\end{table}

\subsection{Automated Liquid Helium Fill System}

The capacity of the volume available for liquid helium in our second-generation Penning-Ioffe trap is roughly 40 L.  Due to the heat load from the presence of 10 vapor-cooled current leads (8 for the Ioffe trap, 2 for the field-boosting solenoid that assists in catching \pbar shots from the AD), our boil-off rate is about 8 L/h.  Since the intention is to remain at 4 K for the entirety of each year's roughly 6-month-long AD beam run, manually adding liquid helium every 5 hours would have severely hindered the ability of our small team to make significant experimental progress.  To overcome this difficulty, a system for automatically filling the trap from a 500 L-capacity storage dewar was installed, reducing the need for manual intervention to one hour every two days.

With a flexible transfer line semi-permanently in place between the transfer dewar and the trap, the autofill system controls the rate of LHe flow into the trap by adjusting the pressure differential between the two as necessary.  The LHe levels in the trap and dewar are monitored using level sensors.  If the level in the trap becomes too low, the control software induces more LHe to flow through the transfer line by reducing the helium pressure in the trap and increasing the pressure in the dewar.  Similarly, if the level rises above a desired limit the control software stops the LHe flow by releasing the pressure from the dewar. 

This system also gives control over the helium exhaust flow rate through each of the 15 ports on the trap.  With prudently selected flow rates, the dewar-trap pressure difference can be balanced well enough to remain within a 5 L-wide control band for the trap LHe level.  The flow controllers incorporated in this system also allow most of the helium exhaust to be forced to pass through the appropriate VCLs during Ioffe trap operation.  This is essential to prevent the VCLs from overheating while they pass up to 680 A of current to the Ioffe trap coils.



\section{Ioffe Trap Operation and Performance}
\label{sec:Operation}

Over the course of several years of operation, the second-generation ATRAP Ioffe trap has been operating close to its design specifications.  This operational performance starts with the ability to energize the trap much more quickly than was possible with the first-generation Ioffe trap.  Multiple diagnostic systems were implemented to ensure that the currents were behaving as expected and that the magnetic fields generated match those calculated based on the measured currents.  Our quench-detection system was tested prior to each operational period and was demonstrated to successfully detect and appropriately respond to quenches, and our rapid-switching IGBT circuit (Fig.~\ref{fig:FullCircuit}) drained the energy from the magnet coils before the quench could proceed far enough to cause damage.



\subsection{Energization}

As discussed in section \ref{sec:CTRAP}, the second-generation Ioffe trap was designed to have a lower inductance to enable faster energization and de-energization than was possible with ATRAP's first-generation trap.  Once the IGBT is past the threshold voltage that allows current to flow, the overall behavior of the circuit can be modelled using the simplified circuit shown in Fig.~\ref{fig:EffectiveCircuitCharging}.  

\begin{figure}
	\centering
	\includegraphics*[width=\columnwidth]{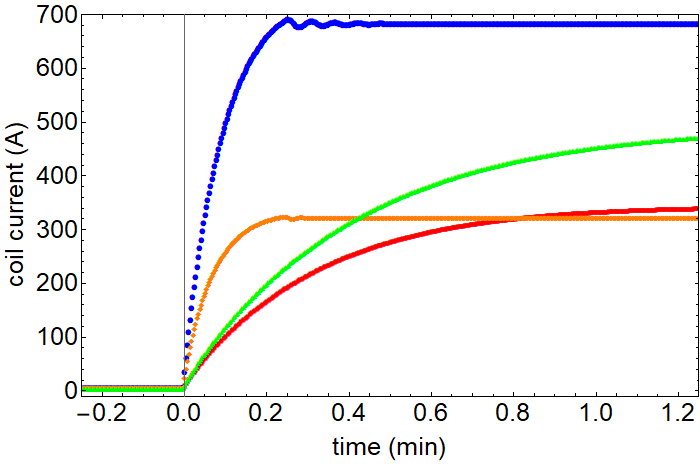}
	\caption{Energization of the individual Ioffe coils, overlaid.}  
	\label{fig:CoilRamps}
\end{figure}

The charging procedure operated under the assumption that a smooth ramp to the desired current would be less likely to result in a quench.  So, to energize a coil the power supply voltage was always raised to just over the threshold at which current began to flow.  Once it was at that point the simplification in Fig.~\ref{fig:EffectiveCircuitCharging} is valid since the IGBT begins to act like a short, aside from reducing the voltage drop over the inductor and resistor by the IGBT threshold voltage.  Increasing the power supply voltage in one step to that required for the desired current then results in a smooth exponential ramp up to the operating current.  Example ramps of each individual coil can be seen in Fig.~\ref{fig:CoilRamps}.

Given the features of the charging circuit, the current should change with time as
\begin{equation}
I = \frac{V_{PS}-V_{th}}{r}\left(1-e^{-tr/L}\right),
\label{eq:Charging}
\end{equation}
where $r$ is the resistance of the leads in series with the coil, $L$ is the inductance of the coil, $V_{PS}$ is the power supply output voltage, and $V_{th}$ is the effective IGBT threshold voltage. 

\begin{figure}
	\centering
	\includegraphics*[width=\columnwidth]{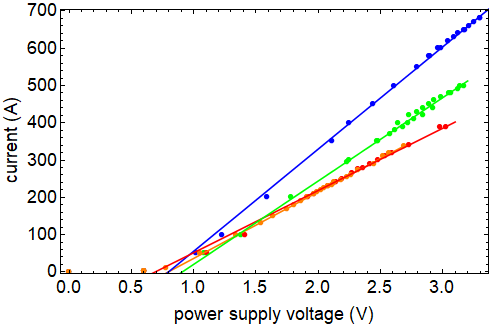}
		\caption{The measured supply voltage and output current show the IGBT threshold and the series resistance for octupole (blue), quadrupole (green), pinch (red) and bucking (orange) coil circuits.}  
	\label{fig:ThresholdAndResistance}
\end{figure}

For the purpose of predicting the time dependence of ramps to high current, the effective IGBT threshold is taken to be the point at which a linear fit to the power-supply voltage vs. steady-state output current intersects the abscissa (see Fig.~\ref{fig:ThresholdAndResistance}).  Since the IGBT output is not linear at low voltages, the low-current data is excluded from the fit that gives the effective threshold values in Table~\ref{table:ThresholdResistance}.  The lower voltages at which the no-current to small-current transition occurs are shown in a separate column.

One can see ringing in the octupole and bucking coil ramp data shown in Fig.~\ref{fig:CoilRamps}.  This ringing is also present in the typical pinch and quadrupole ramps, but not visible in Fig.~\ref{fig:CoilRamps}.  The ringing is due to setting the power supply voltage to a value higher than is necessary to reach the set current.  Once the power supply reaches the set output current, it overshoots somewhat and switches to constant-current mode whereupon the output rings down to the set value.  

There are two advantages to setting the voltage to overshoot the set current.  The first is that the coil can reach a steady output at the set current more quickly than it would through an exponential ramp that has just enough voltage to reach the desired current.  The second is that it allows the power supply's internal control circuit to keep the output current constant without intervention from our DAQ computer.  As these coils stay energized, the high-current leads begin to warm up, increasing the resistance and thus the voltage necessary for a constant current.  Since the power supply voltage during the ramp-up is set to be higher than initially needed, the power supply internal control circuit can adjust the output voltage upwards to compensate.

\begin{figure}
	\centering
	\includegraphics*[width=\columnwidth]{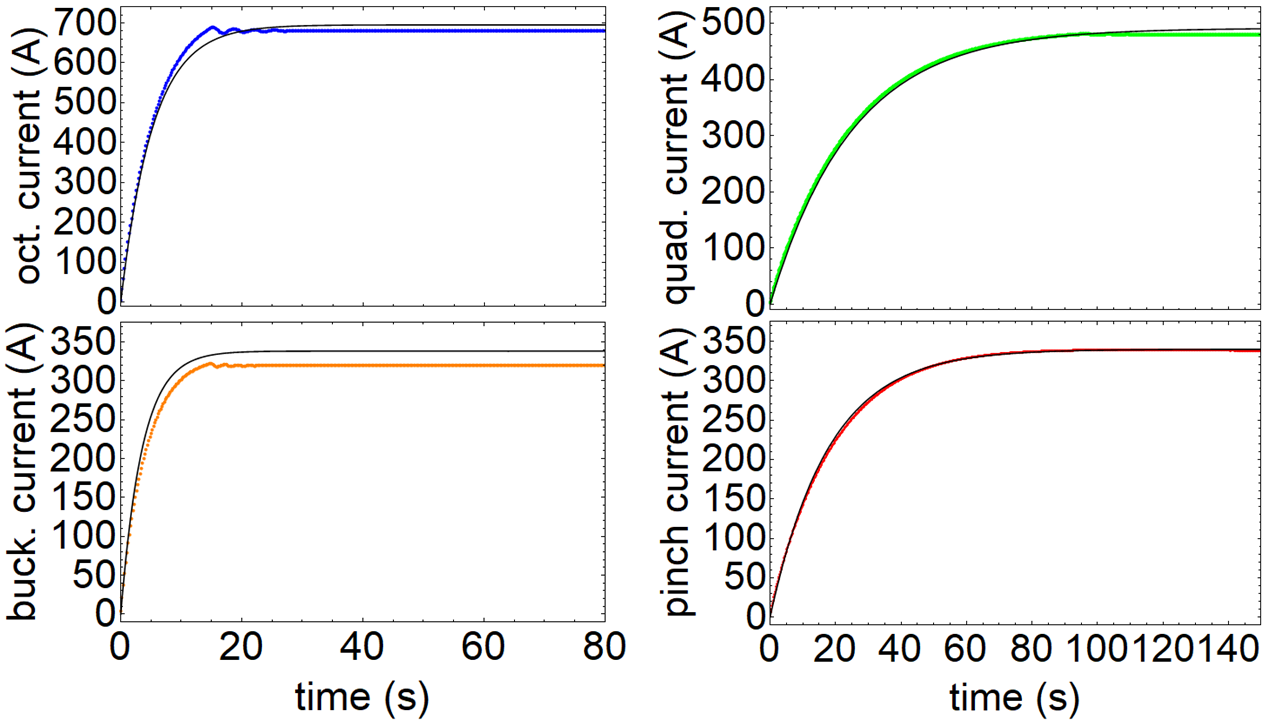}
	\caption{Energization of each of the Ioffe coils.  The solid black curves are the prediction from Eq.~\ref{eq:Charging} using the values in Table~\ref{table:ThresholdResistance}, and the colored points are measured currents.}  
	\label{fig:CoilRampComparisons}
\end{figure}

Fig.~\ref{fig:CoilRampComparisons} shows reasonable agreement between the prediction of Eq.~\ref{eq:Charging} (using the data from Table~\ref{table:ThresholdResistance}) and the coil ramps.  One can see that the agreement with the higher inductance coils is better than for the low-inductance ones.  These ramp predictions are very sensitive to the value of the lead resistance $r$ -- if $r$ = 0.0034 $\Omega$ is used for the octupole instead of the table value of 0.0036 $\Omega$, the predicted plot would very closely match the ramp data.

\begin{figure}
	\centering
	\includegraphics*[width=\columnwidth]{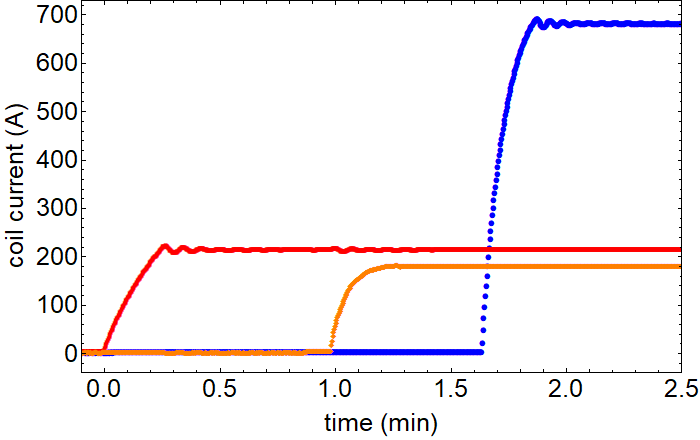}
	\caption{Coil currents during the automated ramp-up routine for a full-depth octupole trap.}  
	\label{fig:TrapRamp}
\end{figure}

Fig.~\ref{fig:TrapRamp} shows the coil currents during an energization of the full octupole trap (the set of currents in Table~\ref{table:DeepestTrap}).  By energizing the coils in the order shown, the aim was to minimize the disturbance of the \pbar and \pos plasmas in the Penning trap, since the coils that have primarily axial fields were turned on before the primarily-radial octupole field was ramped up.  One can also see that the mutual inductance between the pinch and bucking coils results in the bucking ramp-up affecting the current running through the pinch coil.  The sequence used for Fig.~\ref{fig:TrapRamp} results in a fully energized octupole trap with steady currents in just over two minutes.

\begin{table}
	\caption{Thresholds and series resistances of Ioffe circuits}
	\resizebox{\columnwidth}{!}{
	\begin{tabular}{rccc}
	coil	& eff. thresh., $V_{th}$ (V) & curr. thresh. (V) & ~~r ($\Omega$)~~ \\ 
		\hline  \noalign{\vskip 2mm} 
		octupole    &  0.80 & 0.51 & 0.0036\\ 
		quadrupole  &  0.91 & 0.48 & 0.0045\\ 
		pinch       &   0.69 & 0.46 & 0.0060\\ 
		bucking     &   0.80 & 0.46 & 0.0056
	\end{tabular}}
	\label{table:ThresholdResistance}  
\end{table}

\begin{table}
	\caption{Ramp up time constants for the Ioffe coils.}
	\begin{tabular}{rc}
		& ~~~~~L/r in s~~~~~ \\
		\hline
		octupole & 5.1 \\ 
		quadrupole & 24 \\
		pinch & 18 \\ 
		bucking & 3.7 \\ 
	\end{tabular}
	\label{table:RampTimeConstants}
\end{table}

\subsection{Diagnostic Measurements}
\label{sec:Diagnostics}

The bias field, a field-boosting solenoid used only during \pbar loading, and the pinch coil must produce fields that point in the same direction along the central z axis, while the bucking coil produces a field in the opposite direction.  It is important to confirm that the various coils are actually energized when intended and that the magnetic trapping field disappears at the same rate as the coil currents when we dump the trap for \Hbar detection.

Two independent measurement systems monitor the behavior of our trap coils, and an additional one monitors all of our magnets.  The first is the system monitoring the voltage taps, discussed in Section \ref{sec:qpCircuit}, that allows the voltage over the individual coils of the Ioffe trap to be monitored.  The second is a set of current sensors for those coils, and the third is a set of Hall probes for measuring the magnetic field in the vicinity of the Ioffe trap.  These latter two systems are discussed in more detail below.

\subsubsection{Current Sensors}

A Hall-effect current sensor (Tamura L34S1T0D15) is placed at the current return of each power supply in order to have an independent measure of the current running through the Ioffe magnet coils.  These sensors can measure currents up to 1000 A and have a response time of less than 5 $\mu$s, so they could be used to directly measure the current during fast dumps.  Data from these sensors are used in Figs.~\ref{fig:CoilRamps}, ~\ref{fig:CoilRampComparisons}, and ~\ref{fig:TrapRamp}.

\subsubsection{Magnetic Field Sensors}

\begin{figure}
	\centering
    \includegraphics*[width=\columnwidth]{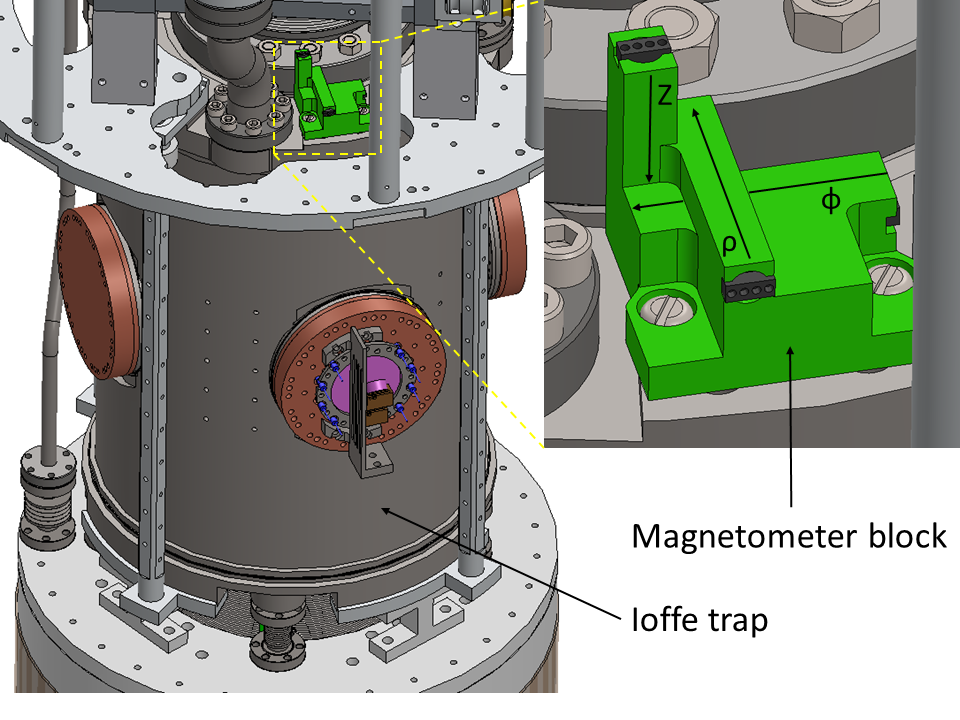}
	\caption{Location and axes of the Hall probes used to check and verify the Ioffe trap fields.}
	\label{fig:Magnetometers}
\end{figure}

Three Hall probes (Cryomagnetics HSP-A) are mounted to the top of the Ioffe trap as shown in Fig.~\ref{fig:Magnetometers}, arranged such that they measure the field components in three orthogonal directions.  Potted with Stycast 2850FT black epoxy into pockets in a custom G10 block (Fig.~\ref{fig:Magnetometers} inset), they face directions that are close enough to cylindrical coordinate axes that those directions ($\rho$, $\phi$, z) are used as the naming convention for the individual probes.

These probes are wired and positioned such that the signal for a magnetic field directed upwards along the trap axis would generate a positive voltage from the $z$-probe, a field directed radially outward would give a positive voltage in the $\rho$-probe, and a field directed to the right in Fig.~\ref{fig:Magnetometers} would give a positive voltage in the $\phi$-probe.  The sensitivities of the $(\rho,\phi,z)$ probes are (27.2, 38.5, 31.4) mV/T with a 50 mA control current and are designed to operate at 4.2 K.  Table~\ref{table:Magnetometers} shows the positions of each probe with respect to the center of the Ioffe coils, along with their response to fields generated by each individual coil.  These positions are from a coordinate system defined by the CNC data for the coil windings provided by AML, where the $z$-axis is opposite our trap orientation.

In addition to monitoring the fields produced by the Ioffe coils, these probes are also useful as a check on the directions of the fields in the Penning-trap solenoid and in the field-boosting solenoid used for enhancing our \pbar catching efficiency.  For example, with no current in the Ioffe coils or field-boosting solenoid, the $z$-probe was at -32.4 mV, indicating that the Penning-trap field was pointing downwards as desired.  Changes in the probe readings could then be monitored as current is added to other coils to ensure that the pinch coil and field-boosting solenoid would add to the Penning field, and that the bucking coil would subtract.

With a 4-channel analog input module (NI 9239) measuring the probe voltages at a rate of 2 kHz, these probes can be used to measure the changing field during fast current dumps.  The fourth channel monitors the TTL level that indicates the status of the quench protection system (discussed in Section~\ref{sec:qpCircuit}), so the voltage-tap measurement data from that DAQ can be synchronized with the Hall-probe data using the TTL change to indicate when the current dump was initiated.  


\begin{figure}
	\centering
    \includegraphics*[width=\columnwidth]{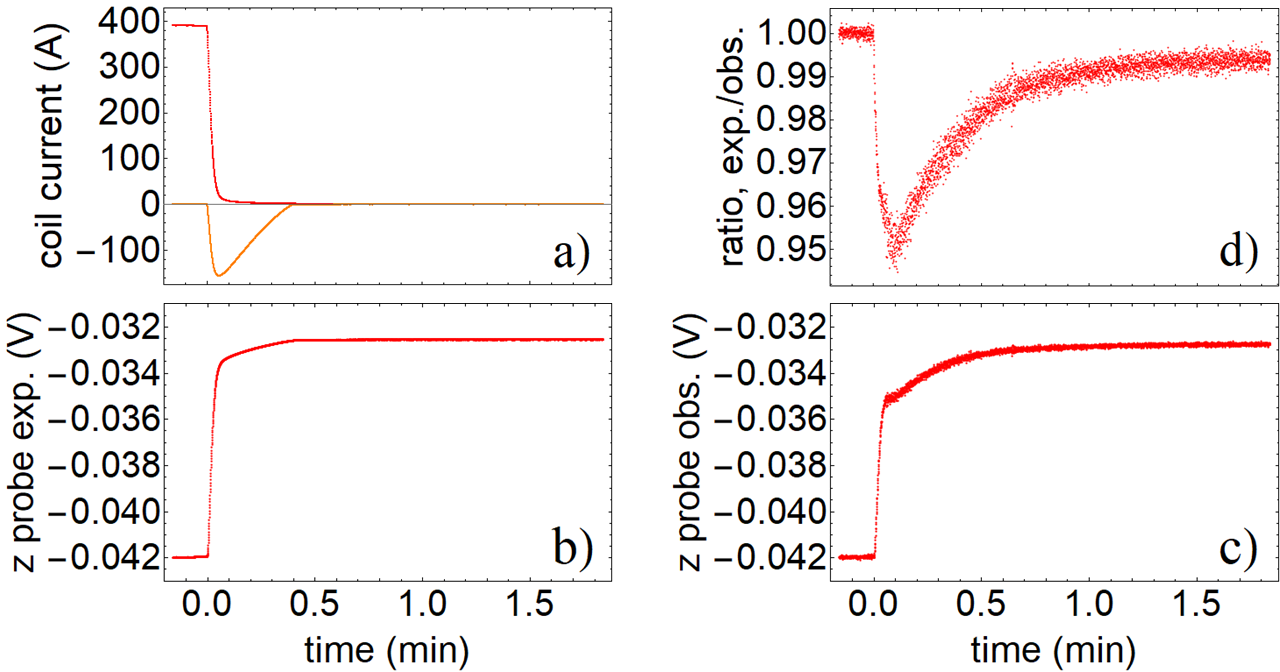}
	\caption{Comparison of expected and measured Hall probe data during a fast dump of the pinch coil: (a) The measured currents in the pinch (red) and bucking (orange) coils (b) The $z$-probe response we would expect from these currents (c) The measured $z$-probe voltages (d) The ratio of the expected voltages to the measured voltages.}
	\label{fig:PinchMagnetometers}
\end{figure}

\begin{figure}
	\centering
    \includegraphics*[width=\columnwidth]{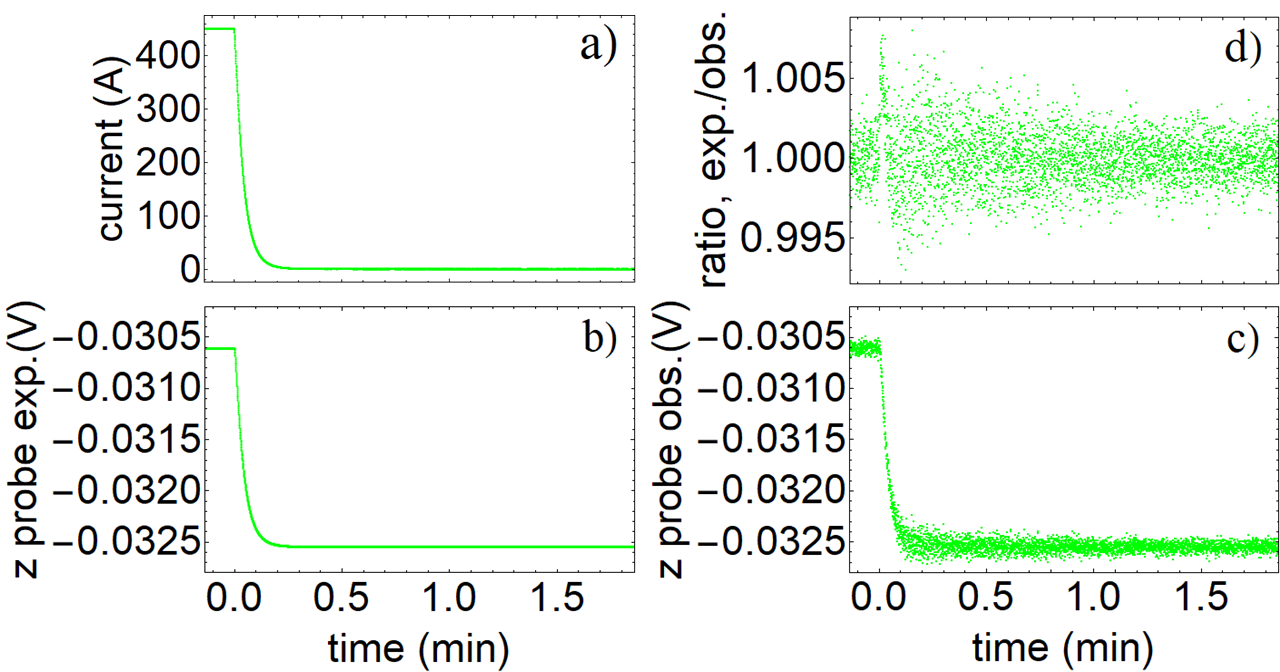}
	\caption{Comparison of expected and measured Hall probe data during a fast dump of the quadrupole coil: (a) The measured currents (b) The $z$-probe response we would expect from these currents (c) The measured $z$-probe voltages (d) The ratio of the expected voltages to the measured voltages.}
	\label{fig:QuadMagnetometers}
\end{figure}

Figs.~\ref{fig:PinchMagnetometers} and \ref{fig:QuadMagnetometers} show some of the Hall probe data obtained during fast dumps of the pinch and quadrupole coils, respectively.  These figures also indicate that the effect of eddy currents caused by dumping the current from the pinch coil is much more noticeable than that from dumping the quadrupole coil. 

Fig.~\ref{fig:PinchMagnetometers}a shows the measured currents during the fast dump of the pinch.  Due to the mutual inductance between the pinch and bucking coils, dumping 390 A from the pinch coil results in a current spike of 156 A in the bucking coil. Given these currents and the sensitivity of the $z$-probe, the expected Hall voltage at each measured current can be calculated (Fig.~\ref{fig:PinchMagnetometers}b). To eliminate effects from uncertainty in the mV/T sensitivity and the value of the control current at the time of measurement, the $z$-probe readings just before and long after the current dump were used to perform a self-calibration of the probes that went into the data in Fig.~\ref{fig:PinchMagnetometers}b.  The measured $z$-probe response during the fast dump of the pinch coil is in Fig.~\ref{fig:PinchMagnetometers}c, and the ratio of the expected values to the measured values is in Fig.~\ref{fig:PinchMagnetometers}d.  This ratio indicates that the eddy current effects in this case are at the 5\% level at the beginning of the dump, and remain at the 1\% level a couple seconds after the current has been completely removed from the coils.

A similar analysis was done for the data in Fig.~\ref{fig:QuadMagnetometers}, where a fast dump of the quadrupole coil was initiated from a starting current of 450 A.  In this case, the mutual inductance with other coils is small enough that the induced currents are negligible.  Comparing the expected and measured $z$-probe readings gives a ratio plot (Fig.~\ref{fig:QuadMagnetometers}d) that deviates from unity only at the very beginning of the dump and by less than 1\%.  This suggests that eddy current effects from dumping the quadrupole coil are small.

\begin{table}
	\caption{Locations with respect to center of the coils and calculated sensitivities (in terms of the coil currents) for the $\rho$, $\phi$ and $z$ probes.}
	\resizebox{\columnwidth}{!}{
	\begin{tabular}{rccccc}
		& ~~$(\rho,\phi,z)$ & octupole & quadrupole & pinch & bucking \\ 
		& (mm, deg, mm) & ($\mu$V/A) & ($\mu$V/A) & ($\mu$V/A) & ($\mu$V/A) \\ 
		\hline  \noalign{\vskip 2mm} 
		$\rho$   & ~~~~$(82,281,-204)$ & -0.144 & -2.26 & -21.6 & +2.31  \\ 
		$\phi$  & ~~~~$(97,287,-196)$ & -0.527 & -1.33 & +6.74 & -0.732  \\ 
		$z$    & ~~~~$(86,286,-187)$ & +0.124 & +4.21 & -29.9 & +5.09  \\ 
	\end{tabular}}
	\label{table:Magnetometers}
\end{table}

\subsection{Quench Detection}

The external protection circuit, described in Section \ref{sec:qpCircuit}, reliably detected quench events and triggered a switch of the IGBT state to drain the current in the coil before it could cause damage to the magnet.  The importance of this protection system was reinforced by the result of a quench that occurred during a development test in which the protection system was disabled.  In that event, one of the superconducting leads to the quadrupole coil melted through.  Since this damage was outside the G10 winding block it was possible to repair it and continue to use the quadrupole coil.

Our quench protection electronics boxes allow the voltage threshold at which a current dump is triggered to be set manually.  While they can trigger on imbalances in the voltage drops over the two VCLs, the two busbars, or the two halves of the coil in the magnet circuit, our standard practice is to set the lowest threshold for the voltage difference between the two magnet halves.  Since the voltage imbalance produced by a quench is most likely to be seen there first, we aim to have the threshold as low as possible, such that the protection circuit will respond as soon as a voltage difference develops that is clearly above the noise level.

\begin{figure}
	\centering
	\includegraphics*[width=\columnwidth]{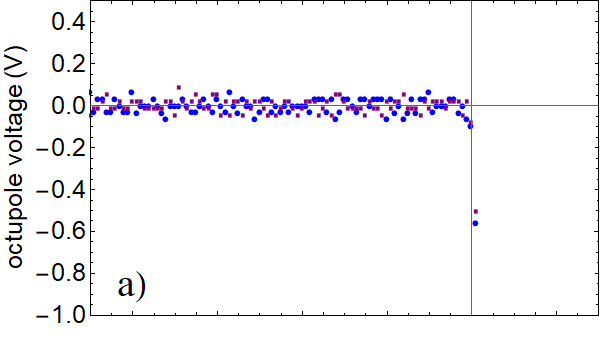}
	\includegraphics*[width=\columnwidth]{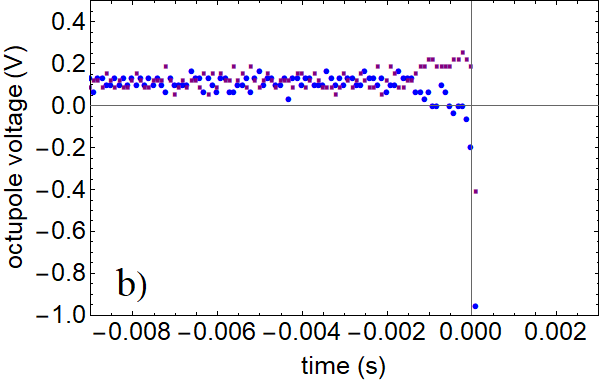}
	\caption{The voltages across halves of the octupole coil just prior to a triggered dump (top) and a detected quench (bottom).  Time zero indicates when the quench protection circuit triggered.  Blue circles represent $V_C - V_D$, and the purple squares are $V_D - V_E$.}  
	\label{fig:DumpVersusQuench}
\end{figure}

While the voltages actually seen by the quench protection circuit are smaller due to the voltage dividers mentioned earlier, voltage tap measurements are presented in Fig.~\ref{fig:DumpVersusQuench} in terms of the voltages seen by the magnet circuit elements.  So, as can be seen in Fig.~\ref{fig:DumpVersusQuench}, the peak-to-peak noise level of the voltage measurements over each half of a coil is about 100 mV.  Our standard threshold setting for the imbalance between magnet halves is 250 mV.  The line at t=0 in the plots in Fig.~\ref{fig:DumpVersusQuench} represents the time at which the quench protection circuit triggered a fast current dump.  In Fig.~\ref{fig:DumpVersusQuench}a the dump was triggered manually, while in Fig.~\ref{fig:DumpVersusQuench}b a quench detection triggered the current dump.  One can see that in the case of a quench the voltage drops over halves of the magnet start to diverge.  The voltage drops over the magnet halves in Fig.~\ref{fig:DumpVersusQuench}b prior to the quench were non-zero because the quench occurred while the current in the coil was being ramped up.  In both cases, once the quench protection circuit is triggered (either manually or through a detected quench) the measured magnet voltages behave in the same manner.

As discussed in more detail in Refs. \cite{ThesisNovitski} and \cite{WilsonSuperconductingMagnets}, during a quench the voltage over the magnet terminals remains relatively unchanged as the resistive voltage drop due to a section of the superconductor becoming normally-conducting is in opposition to the voltage induced by the changing current in the inductive coil.  Thus, as a quench begins, there is little to no change in the measurements from voltage taps C and E, while the voltage at tap D will either increase or decrease depending on which half of the magnet contains the quenching region.  Since the voltage drops over magnet halves are plotted in Fig.~\ref{fig:DumpVersusQuench} ($V_C - V_D$ and $V_D - V_E$), we see that, as a quench begins, these two voltage drops change in opposite directions by approximately equal amounts.

\subsection{Rapid Current Dumps}
\label{section:RapidDumps}

\begin{figure}
	\centering
	\includegraphics*[width=0.6\columnwidth]{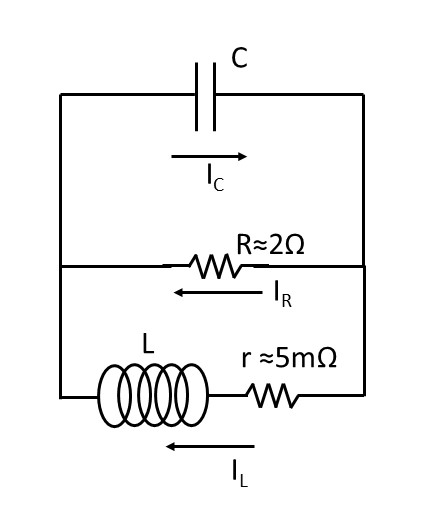}
	\caption{The effective circuit for dumping an individual Ioffe coil.}  
	\label{fig:DumpCircuit}
\end{figure}

Upon detection of a quench, our protection system triggers, rapidly removing the current from the coils to protect them from damage.  When we want to release \Hbar from our trap for detection, it is shut off by manually triggering the same protection system.  In either case, once the protection system opens the IGBT the effective circuit can be drawn as in Fig.~\ref{fig:DumpCircuit}.  This is similar to the effective charging circuit from Fig.~\ref{fig:EffectiveCircuitCharging}, but with the IGBT unit's capacitor in place of the power supply.

Since the resistances, capacitances, and inductances of the components in Fig.~\ref{fig:DumpCircuit} have been measured independently, the circuit can be solved to generate a prediction for the voltages over a Ioffe coil during a fast current dump.  The differential equation for the current through the inductor, $I_L(t)$, in terms of the time constants $\tRC=RC$ and $\tLR=L/R$ is
\begin{equation}
\tRC\tLR \ddot{I}_L + \tLR \dot{I}_L + \left(\frac{r+R}{R}\right)I_L = 0.
\end{equation}
With the initial condition of a constant current $I_0$ running through the inductor, the exact solution can be written as
\begin{equation}
I_L = I_0 \left[ 
\frac{1+s}{2s} e^{-\frac{t(1-s)}{2\tRCs}} - \frac{1-s}{2s} e^{-\frac{t(1+s)}{2\tRCs}} \right],
\label{eq:PredictedIL}
\end{equation}
where
\begin{equation}
s \equiv \sqrt{1-4\left(\frac{r+R}{R}\right)\frac{\tRC}{\tLR}}.
\label{eq:sDef}
\end{equation}

With Eq.~\ref{eq:PredictedIL} giving the form of the current through a Ioffe coil during a rapid dump, the expression for the voltage spike over the coil can be written down from $V = LdI_L/dt$, giving
\begin{equation}
V_{\textrm{coil}} = \frac{I_0R}{s}\left[e^{-\frac{t(1+s)}{2\tRCs}} - e^{-\frac{t(1-s)}{2\tRCs}}\right].
\label{eq:PredictedVL}
\end{equation}
Our quench protection system continuously monitors the voltages across the coils using the voltage taps discussed in Sec.~\ref{sec:qpCircuit}.  As mentioned there, the voltage spike first passes through a cryogenic voltage divider to reduce the amplitude by a factor of five.  A roughly 5 meter long cable made using Alphawire 6383 (ground capacitance of 207 pF/m) carries the tap voltages to the quench protection electronics.  Here they pass through another voltage divider to reduce their amplitude by another factor of twenty, in order to keep them within the operating range of the circuit components.  The voltages that our protection system monitors (and records at a rate of 10 kHz in the event of a quench or triggered dump) are those after the voltage dividers and cabling.  To obtain the voltage observed by the voltage monitors in the quench protection circuit, $V_{\textrm{mon}}$, the differential equation
\begin{equation}
    V_{\textrm{coil}} = 100 V_{\textrm{mon}} + (0.012) \dot{V}_{\textrm{mon}}
\end{equation}
can be solved.  

\begin{figure}
	\centering
	\includegraphics*[width=\columnwidth]{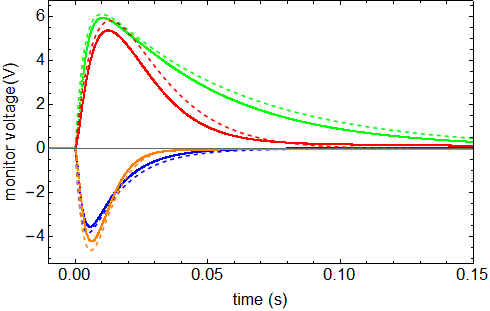}
	\caption{Measured (solid) and predicted (dashed) voltages in the quench protection circuit produced by rapid dumps of the octupole (blue), quadrupole (green), pinch (red) and bucking (orange) coils, respectively.}  
	\label{fig:DumpVoltages}
\end{figure}

Fig.~\ref{fig:DumpVoltages} shows examples of the monitor-voltage spikes for rapid current dumps of each individual Ioffe coil.  The initial currents in each are $I_0 = $400, 350, 200, and 340 amperes for the octupole, quadrupole, pinch, and bucking coil, respectively.  In each case, it is observed that the predicted voltage spike is slightly higher than the measured value.  If, instead of using the measured values of the dump resistor, coil inductance, and IGBT-unit capacitance, they are allowed (along with the capacitance of the cable between the trap and the quench protection system) to be variable parameters in a fit to the octupole data, much better agreement between the curve and the data can be obtained.  The best-fit values for R, L, and C are within about 15\% of the independently measured values, while the cable capacitance in the fit increases by an order of magnitude.  This indicates that there are significant additional stray capacitances in this distributed circuit that aren't accounted for by the model based on Fig.~\ref{fig:DumpCircuit}.

\begin{figure}
	\centering
	\includegraphics*[width=\columnwidth]{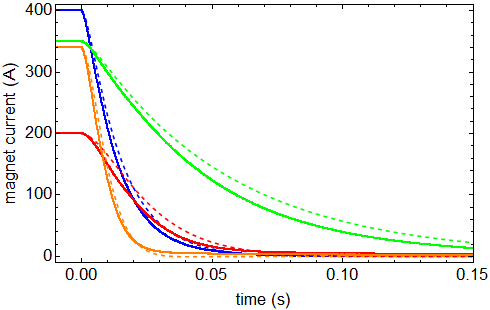}
		\caption{Current in the octupole (blue), quadrupole (green), pinch (red) and bucking (orange) coils as measured by the current sensors (points) together with the current predicted in Eq.~\ref{eq:PredictedIL} (dashed).} 
	\label{fig:Currents}
\end{figure}

As introduced in Sec.~\ref{sec:Diagnostics}, a set of Hall-effect current sensors provide an independent means of monitoring the behavior of the Ioffe coils during a rapid dump.  These devices are used to record the current during the rapid dumps of Fig.~\ref{fig:DumpVoltages} at the same (10 kHz) rate as the voltage data.  Fig.~\ref{fig:Currents} shows this data along with the predicted curve from Eq.~\ref{eq:PredictedIL}.  As in Fig.~\ref{fig:DumpVoltages}, the predicted curve shows reasonably good qualitative agreement with the measurements, with the remaining discrepancy likely reflecting features of the physical circuit that aren't included in the model, such as distributed capacitance.

While our model of the circuit during a rapid dump clearly neglects features that have a noticeable effect on the shapes of the voltage and current curves, it's equally clear that the model includes the primary contributions to those curves.  This model could then be used, for example, to guide an effort to fine-tune the dump resistance values to optimize the current decay time while keeping the height of the voltage spike below a chosen limit.  The qualitative agreement between the model and the measurements adds to our confidence that the predicted effect of changing the dump resistance will match the effect that would be observed in the actual circuit.



\section{Signals of Trapped \pos, \pbar, and \Hbar}
\label{sec:Usefulness}

To illustrate the usefulness of the Penning-Ioffe trap two examples are used.  For charged particles, images of antiproton and positron plasmas are shown.  Evidence of the robust trapping of neutral antihydrogen atoms is also presented.  

\subsection{Antimatter Plasmas}
\label{sec:PositronPlasmas}

A plasma imaging system inside the cryogenic vacuum space of the Penning-Ioffe trap characterizes the electron, positron, and antiproton plasmas used during our anithydrogen production procedure.  By running through a sequence of plasma preparation steps \cite{ThesisNottet} and then ejecting the plasma up to the imaging system, it could be confirmed that we were reproducibly preparing plasmas with appropriate radii at each stage of the procedure.

\begin{figure}
	\centering
    \includegraphics*[width=\columnwidth]{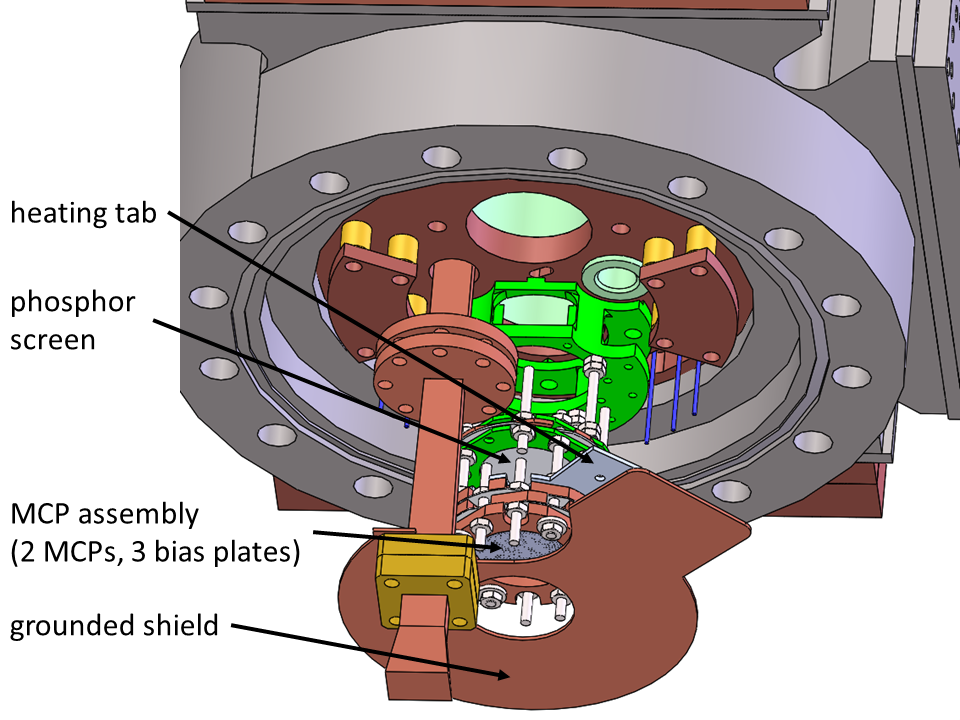}
	\caption{The plasma imaging system}
	\label{fig:PlasmaSystem}
\end{figure}

Fig.~\ref{fig:PlasmaSystem} shows the components of this system, which were mounted to the bottom of a translation stage that contains the upper boundary of the Penning trap volume and allows multiple devices, viewports, and small through-holes to be moved to the center of the trap axis. The grounded copper shield, located 9 cm above the topmost Penning-trap electrode, helps protect the insulating components of the imaging system from charging up during particle loading and shields the Penning-trap electrodes from the MCP bias voltages.  Our MCP assembly consists of two Hamamatsu F1094-01 MCPs and three bias plates to allow us to control the voltage across each MCP individually.  The base of the MCP assembly includes an anodized aluminum tab located below a viewport through which a heating laser can be sent, as repeated firing of the MCPs at their 12 K measured equilibrium temperature would be expected to deplete the electrons in the channels.  A plasma that impacts the bottom MCP results in a similarly-shaped cloud of electrons accelerated towards the phosphor screen (Kimball Physics PHOS-UP22GL).  

\begin{figure}
    \centering
    \includegraphics[width=\columnwidth]{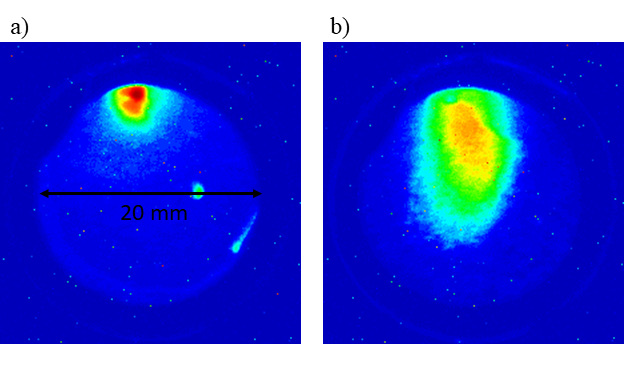}
    \caption{a) An example image of a positron plasma, with the 20 mm diameter of the visible surface of the phosphor screen indicated.  The movable stage the imaging system was mounted to did not have sufficient range of motion to center the imaging system on the trap axis. b) An example image of a less-compressed antiproton plasma.}
    \label{fig:PlasmaImages}
\end{figure}

The light from the phosphor is recorded using a camera located out of vacuum at a distance of about 1 meter from the phosphor screen. Fig.~\ref{fig:PlasmaImages} shows two example images, one for an antiproton plasma and one for a positron plasma. The effective active area of the phosphor is 20 mm diameter due to the mounting hardware.  The edge of the 1-inch diameter MgF$_2$ window just above the phosphor screen is visible in the full images, so that diameter is used to generate a pixels per mm calibration.  Typical plasma radii used for antihydrogen production trials were 4 mm for antiprotons and 1 mm for positrons.

\subsection{Trapped Antihydrogen}
\label{sec:TrappedAntihydrogen}

Towards the end of the 2018 antiproton beam run at CERN's antiproton decelerator facility, we developed a procedure for repeatably preparing antiproton and positron plasmas in a nested well for \Hbar production.  In 9 trials performed identically, average annihilation signals above background were observed, indicating the successful production and confinement of antihydrogen atoms in our second-generation Penning-Ioffe trap.

\begin{figure}
    \centering
    \includegraphics[width=\columnwidth]{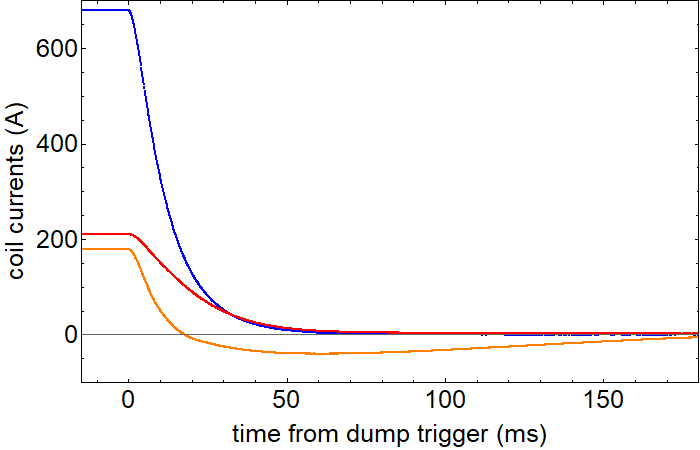}
    \caption{Measured currents during a fast dump of the octupole trap, with the octupole coil current in blue, the pinch current in red, and the bucking current in orange.}
    \label{fig:OctTrapDumpCurrents}
\end{figure}

\begin{figure}
    \centering
    \includegraphics[width=\columnwidth]{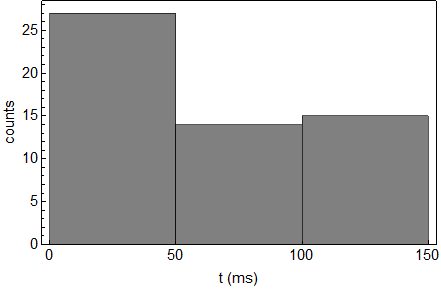}
    \includegraphics[width=\columnwidth]{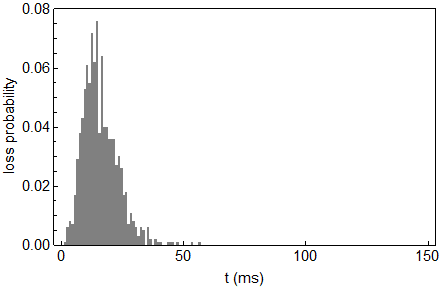}
    \caption{\Hbar annihilation data (top) and simulations (bottom), with t=0 corresponding to the initiation of a fast dump of the Ioffe trap.  Virtually all trapped \Hbar should annihilate in the first 50 ms, so counts during the next two time segments give the background rate.}
    \label{fig:HbarObserved}
\end{figure}

In each of these trials the detection method was destructive: after running through the procedure to induce the \pbar and \pos to interact and form \Hbar through three-body recombination while the Ioffe trap was energized, a 5 V/cm electric field was applied to remove any remaining charged particles, then a trigger was sent to the quench protection boxes to dump the current from the Ioffe trap.  This released any trapped \Hbar, which annihilate on the electrode stack producing pions that can be observed by our particle detectors \cite{ThesisZhang}. Given the coil currents measured during a fast dump of the octupole trap (Fig.~\ref{fig:OctTrapDumpCurrents}), it can be determined that after 35 ms there is no radial confinement and the axial confinement has decreased by 94\%, and after 50 ms essentially no \Hbar confinement remains.  This is consistent with the simulated \Hbar release times in the bottom plot of Fig.~\ref{fig:HbarObserved}, and motivates the 50 ms time bins in the annihilation data from nine trials shown in the upper plot of Fig.~\ref{fig:HbarObserved}.  The excess of counts in the first 50 ms of data after sending the dump trigger corresponds to $5\pm2$ \Hbar per trial.



\section{Conclusion}
The ATRAP collaboration developed and utilized two implementations of a combined Penning-Ioffe trap, both of which featured radial access ports to allow trapped particles to be addressed with lasers or microwaves travelling perpendicular to the Penning-trap magnetic field direction.  These traps both reached significant milestones, showing evidence of trapped antihydrogen, as well as, in the case of the first Penning-Ioffe trap, demonstrating a novel laser-controlled, two-step charge exchange procedure for producing \Hbar.

ATRAP's second Penning-Ioffe trap was built with two sets of coils for generating a radial field gradient, allowing for the neutral particle trap to be energized in either a quadrupole or octupole symmetry.  Its low-inductance, high-current coils allowed for rapid ramp-up and de-energization, which is advantageous for \Hbar production and destructive detection.  The high-current feedthroughs required for operation cause the rate of liquid helium usage to be significantly higher than that of ATRAP's first Penning-Ioffe trap, but was compensated for by implementing a reliable automated LHe filling system.

A suite of voltage, current, and magnetic field diagnostics provides means of monitoring the performance of the Ioffe trap coils with sufficient redundancy to confirm that the trap depth and turn-on/turn-off times were all reaching the design goals.  In addition, the plasma imaging system aided in the development of a reproducible procedure for preparing \pbar and \pos plasmas for \Hbar production.  The resulting evidence of \Hbar detection demonstrates the potential of this apparatus for precision spectroscopy of antihydrogen atoms.

	

\section{Acknowledgments}

We are grateful to Bruce Gold at Joining Technologies for his intricate electron beam welding of the titanium enclosure for the Ioffe coils.  We'd like to recognize the following former members of the ATRAP collaboration for their contributions to early development tests of the second-generation Ioffe trap: Rita Kalra, Martin Kossick, and Stephan Malbrunot (n\'{e} Ettenauer).  M. Kossick also contributed to early design work for the VCLs, and S. Malbrunot contributed to the vacuum enclosure design.  We thank Phil Richerme for his early work on the quench protection system and large-diameter electrodes. This work was supported by the NSF and by the AFOSR.



\bibliographystyle{prsty_gg}
\bibliography{ggrefs2018}

\end{document}